\documentclass[12pt]{iopart}
\pdfoutput=1
\usepackage{iopams}
\usepackage{graphicx}	
\usepackage{subfig}		

\newcommand{\Bt}{$^{10}$B}
\newcommand{\Ls}{$^{6}$Li}
\newcommand{\Het}{$^3$He}
\newcommand{\Uttf}{$^{235}$U}

\begin{document}
\newpage

\title[]{Precision determination of absolute neutron flux}

\author{A.T.~Yue$^1$, E.S.~Anderson$^{2}$, M.S.~Dewey$^1$, D.M.~Gilliam$^1$, G.L.~Greene$^{3,4}$, A.B.~Laptev$^5$, J.S.~Nico$^1$, W.M.~Snow$^{2}$}

\address{$^1$ National Institute of Standards and Technology, Gaithersburg, MD 20899 USA}
\address{$^2$ Indiana University, Bloomington, Indiana 47408, USA}
\address{$^3$ University of Tennessee, Knoxville, Tennessee 37996, USA}
\address{$^4$ Oak Ridge National Laboratory, Oak Ridge, Tennessee 37831, USA}
\address{$^5$ Los Alamos National Laboratory, Los Alamos, New Mexico 87545, USA}

\ead{maynard.dewey@nist.gov}

\begin{abstract}

A technique for establishing the total neutron rate of a highly-collimated monochromatic cold neutron beam was demonstrated using a method of an alpha-gamma counter. The method involves only the counting of measured rates and is independent of neutron cross sections, decay chain branching ratios, and neutron beam energy. For the measurement, a target of  \Bt-enriched boron carbide totally absorbed the neutrons in a monochromatic beam, and the rate of absorbed neutrons was determined by counting 478\,keV gamma rays from neutron capture on \Bt\ with calibrated high-purity germanium detectors.  A second measurement based on Bragg diffraction from a perfect silicon crystal was performed to determine the mean de Broglie wavelength of the beam to a precision of 0.024\,\%.  With these measurements, the detection efficiency of a neutron monitor based on neutron absorption on ${}^{6}$Li was determined to an overall uncertainty of 0.058\,\%. We discuss the principle of the alpha-gamma method and present details of how the measurement was performed including the systematic effects. We also describe how this method may be used for applications in neutron dosimetry and metrology, fundamental neutron physics, and neutron cross section measurements.

\end{abstract}

\vspace{2pc}

\maketitle

\section{Introduction}
\label{sec:intro}

Accurate counting of slow neutrons is required in a variety of applications. The nuclear power industry uses neutrons to monitor reactor power and to perform diagnostics on reactor operation~\cite{Shultis2007}. Certain medical applications require a good knowledge of neutron fluence to determine doses to patients, particularly boron neutron capture therapy treatments~\cite{Harling1995,Raaijmakers1996,Tanner1999}. It is important in many experiments in neutron scattering and fundamental neutron physics to know the neutron rate to understand performance of instruments or the feasibility of a measurement. The neutron rate may also be an integral component in determining a physical quantity.

There exist many techniques for the detection of slow neutrons~\cite{Beckurts1964,Knoll2000,Williams2011}. In the majority of these techniques, a neutron is incident upon an isotope that has a large reaction cross section and produces energetic charged particles that can be detected through several methods. Typical reaction products are protons, alpha particles, and fission fragments. Detectors that utilize prompt gamma rays exist but are less common because of the long interaction length for gamma rays of the MeV energies typically produced in neutron capture. Some of the more common isotopes used in neutron detectors are \Het, \Ls, \Bt, and \Uttf. The capture material is typically contained in an environment that produces detectable ionization from one or more charged particles resulting from the neutron reaction. Commonly used detectors are \Het\ and BF$_3$ proportional counters, doped scintillators (\Ls,\Bt, Gd), and \Uttf\ fission chambers. Another neutron detection technique, called activation analysis, uses incident neutrons on an isotope with a large absorption cross section to induce subsequent radioactivity amenable to detection~\cite{Greenberg2011,deSoete1972}. Common isotopes used in activation analysis are $^{164}$Dy and $^{197}$Au. The growing field of neutron imaging has developed a wide variety of detection schemes~\cite{Fraser1995,Brenizer2013}, but these researchers are typically not concerned with the absolute rates.

For the majority of applications, accurate neutron fluence is important, but achieving a high precision is not a critical component of the measurement. Broadly speaking, the precision that has been achieved by these counting methods is not significantly better than 1\%, and for many methods it can be much poorer. The agreement resulting from an international key comparison of thermal neutron fluence measurements involving national metrology laboratories is about 3\,\%~\cite{Nolte2015}. A comparable precision exists in international comparisons for measurements of fast neutrons~\cite{Gressier2014}.  A thorough discussion of the status and methods of neutron metrology is found in Ref~\cite{Thomas2011}.

For some applications, there is a strong need to improve the ultimate precision to the level of 0.1\% or better. A notable example is measuring the neutron lifetime using a cold neutron beam. In that method, both the decay protons and beam neutrons must be counted with high accuracy~\cite{Wietfeldt2011}. The limiting systematic uncertainty has historically been the determination of the absolute neutron flux~\cite{Byrne96,Dewey2003,Nico05}, and thus, improving the ultimate precision is critical to improving the neutron lifetime.  A precise neutron flux measurement technique can also be used to improve measurements of important neutron cross section standards (e.g., ${}^{6}$Li, \Bt, and ${}^{235}$U) at near thermal energies~\cite{Dewey08}. In addition, it can be used in an improved measurement of the emission rate of NBS-1, a Ra-Be photo-neutron source that is the national standard in the US~\cite{Adams04,Gilliam08}. These specific applications are discussed in more detail in Section~\ref{sec:summary}.

In this paper, we report the development of a neutron flux monitor and a totally absorbing neutron detector that were used to measure the flux of a monochromatic neutron beam to an overall uncertainty of 0.058\,\%. In the remainder of this section, we briefly discuss other methods for precision neutron measurements. Section~\ref{sec:AGD} presents the concept of the Alpha-Gamma measurement and details of the apparatus, and Section~\ref{sec:FM} has a similar discussion on the flux monitor. The data acquisition system and the analysis methods are discussed in Section~\ref{sec:DAQ}. Section~\ref{sec:lambdasys} covers the measurement of the mean wavelength of the neutron beam. The corrections to the measured neutron flux and the associated systematic uncertainties are addressed in detail in Section~\ref{sec:syst}. Lastly, the results are summarized and future prospects are discussed in Sections~\ref{sec:results} and \ref{sec:summary}.


\subsection{Terminology}

There is some terminology used in this paper that should be clarified at the outset to avoid confusion. The goal of this work is to determine the entire neutron content of a beam per unit time. The particle content of a neutron beam may be characterized in different ways, and therefore, it is important to be clear on the quantity that is being measured. We follow the definitions given in the article entitled ``Fundamental Quantities and Units for Ionizing Radiation"~\cite{Seltzer2011} and reiterate them here. The flux $\dot{N}$ is the quotient of d$N$ by d$t$, where d$N$ is the increment of the particle number in the time interval d$t$, and it has units of s$^{-1}$. By design, the cross-sectional area of all the deposits and targets in this work were larger than that of the neutron beam itself, and thus their areas are not a relevant parameter. When discussing measured particle counting rates, we often simply refer to them as rates.

For many types of measurement, the area is a critical parameter, and the units of fluence or fluence rate are appropriate. The fluence $\Phi$ is the quotient of d$N$ by d$a$, where d$N$ is the number of neutrons incident on a sphere of cross-sectional area d$a$. Fluence has units of m$^{-2}$. It follows that fluence rate is the quotient of d$\Phi$ by d$t$, where d$\Phi$ is the change in a time interval $t$. Fluence rate has units of m$^{-2}$\,s$^{-1}$. When mentioning uncertainties, we are referring to the standard deviation with k=1, or 68\,\% confidence level.

In addition, the thickness of deposits and targets relative to their neutron interactions is an important quantity and falls into two categories that are often referenced by their colloquial names.  The terms  ``black" and ``thick" are used interchangeably throughout and refer to a deposit or target with an areal density such that it effectively absorbs the entire neutron beam.  The term ``thin'' refers to a deposit or target with an areal density that negligibly attenuates the neutron beam. In some instances, the numerical value of an areal density may be given, and in others, the colloquial term is more informative.

Lastly, the detection of alpha particles was done using two types of silicon charged-particle detector. They were either ion-implanted detectors or surface barrier detectors and were manufactured by either Canberra Industries or Ortec (Ametek)\footnote{Certain commercial equipment, instruments, or materials are identified in this paper in order to specify the experimental procedure adequately. Such identification is not intended to imply recommendation or endorsement by the National Institute of Standards and Technology, nor is it intended to imply that the materials or equipment identified are necessarily the best available for the purpose.}. As the distinction between the two detectors is not relevant to any of the measurements discussed here, we refer to them as silicon detectors or charged particle detectors.

\subsection{Overview of precision absolute neutron flux measurements}

There are two mature experimental strategies for determining the absolute neutron rate of a neutron beam at a level of precision below 0.1\,\%. One is the method of alpha-gamma counting, the topic of this paper, in which a device counts alpha particle and gamma ray events from neutron capture to calibrate the rate of gamma-ray production from a totally absorbing target of \Bt~\cite{Lamaze88}. The second method uses an electrical substitution radiometer to determine the power delivered by neutron absorption in totally absorbing targets at cryogenic temperatures~\cite{Robertson86}. We briefly review that method in this section. There are at least two other methods under development that could reach comparable precision using neutron capture on \Het\ to determine the neutron rate~\cite{Wietfeldt09,Nagakura2017}.


The neutron radiometer operates as an absolute neutron detector by measuring the power produced by neutrons absorbed in a target cooled to cryogenic temperatures.  The power is measured with an electrical substitution radiometer, in which the power delivered by radiation absorbed in a target can be compared to an equivalent amount of electrical power.  The target is coupled to a heatsink through a weak thermal link.  The heatsink is kept at a constant temperature difference with respect to the target, and the power required to maintain the temperature difference is monitored.  The heat generated by reaction products from the absorption of the neutron beam can then be determined from the difference in electrical power delivered with the beam on and off.

The radiometer target material must be chosen carefully.  The ideal target is composed of a material that is totally absorbing to neutrons and produces reaction products that contribute a known and measurable amount of heat absorbed in the bulk of the target. Two isotopes, \Ls\ and \Het, were envisioned as target materials in the original proposal. \Ls\ is a good candidate because of its large neutron absorption cross section, high and precisely-known Q-value of ($4.78293 \pm 0.00047$)\,MeV, and readily absorbed reaction products that do not produce gamma-rays.  A target of pure \Ls\ is not practical, and at low temperatures \Ls\ undergoes a first-order phase transition; therefore, a transformation-inhibiting material must be added to make a viable target.  This introduces additional absorption and scattering mechanisms to the target due to the addition of another element.  Furthermore, these alloy targets are solid and polycrystalline, allowing for potentially large, hard to assess coherent scattering effects. Finally, any solid lithium-based target can store a small fraction of energy in metastable lattice defects in an amount that is hard to measure or calculate. For these reasons, \Het\ was also considered as a target.

\Het\ has a significantly lower Q-value of ($0.763763 \pm 0.000004$)\,MeV and a higher heat capacity, making it more technically challenging to perform the power measurement.  Liquid \Het\ does not, however, possess long-term energy storage modes through the mechanism of radiation damage, which can make the more accessible \Ls-based measurements difficult to interpret.  To date, three measurements have been performed with the neutron radiometer using both solid \Ls-based targets and a liquid \Het\ target~\cite{Richardson93,Chowdhuri00,Chowdhuri03,Hansen04}. Only the measurement with the  \Ls Mg target achieved an uncertainty of 0.1\,\%~\cite{Chowdhuri03}; the other experiments were limited by systematic uncertainties or technical problems.

\section{The Alpha-Gamma device}\label{sec:AGD}
\subsection{Principle of measurement}
\label{sec:prinofmeasure}

The Alpha-Gamma method relies on the accurate counting of alpha and gamma radiation emitted from both neutron absorbing materials used as interchangeable targets in a cold neutron beam and radioactive sources used for calibration. The fundamental parameter on which the neutron counting is based is the absolute emission rate of an alpha source, which is determined in measurements performed offline. The Alpha-Gamma device uses that alpha source and the neutron beam to establish its alpha and gamma counting efficiencies, which are used to establish the neutron flux. The method only relies upon measured rates and does not depend upon target cross sections, branching ratios, or knowledge of enrichment fraction. In brief, the neutron flux is determined in the following steps:

\begin{enumerate}
\item determine the absolute activity of an alpha source; 
\item establish the efficiency of an alpha detector inside the Alpha-Gamma apparatus using the calibrated alpha source;
\item transfer the alpha-particle efficiency to a gamma-detector efficiency using the alpha particles and gamma rays produced by neutrons incident on a thin \Bt\ target;
\item finally, determine the neutron flux by counting gamma rays from the interaction of neutrons incident on a thick \Bt\ target.
\end{enumerate}

\begin{figure}[htbp]
  \centering
  \includegraphics[width=1\textwidth]{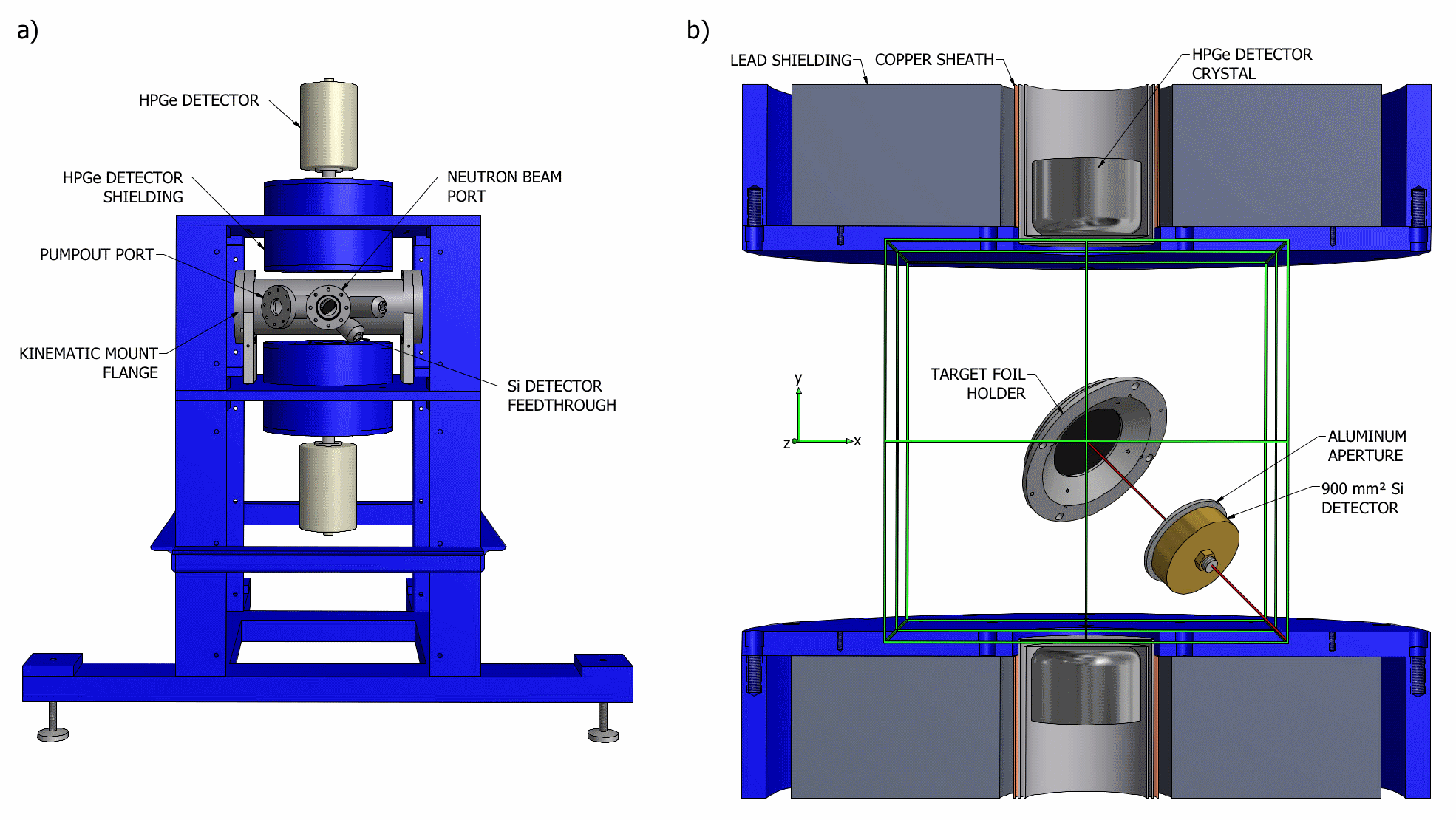}
  \caption{Illustrations of a) the Alpha-Gamma device vacuum system (gray) in its support frame (blue) and b) a section view of the Alpha-Gamma device showing the detection geometry; the neutron beam is incident on the target at the position of the green cross.}
  \label{fig:AGgeo}
\end{figure}

The critical apparatus used to accomplish this experimentally is the Alpha-Gamma device, which uses high-purity germanium (HPGe) detectors and a silicon charged-particle detector to count gamma and alpha radiation, respectively, inside a high vacuum system. The device functions as a black detector by totally absorbing a neutron beam in a 98\,\%-enriched target of ${}^{10}$B${}_{4}$C and counting the emitted 478\,keV gamma-rays in the two HPGe detectors~\cite{Gilliam89}.  An alpha-to-gamma cross calibration procedure determines the number of neutrons absorbed in the ${}^{10}$B${}_{4}$C target per observed 478\,keV gamma. Ultimately, the alpha detector efficiency is established via the well-known alpha source, thus providing the efficiency for neutron counting. Figure~\ref{fig:AGgeo} shows an illustration of the Alpha-Gamma device.

\subsection{Alpha-Gamma apparatus}

The Alpha-Gamma device is composed of three main components: the target holder, the alpha-particle detector, and two HPGe detectors. These components are housed in a cylindrical aluminum high-vacuum system and are supported by a robust steel frame. The frame must be mechanically strong because it also supports lead shielding that surrounds the HPGe detectors to reduce their ambient gamma and neutron backgrounds. Lead bricks fill the empty space surrounding the vacuum chamber, and eight composite panels made of borated rubber sheets captured between thin steel sheet metal are bolted to the outside of the steel frame to reduce neutron backgrounds.  The gamma detectors are mounted into a steel protrusion on the top and bottom of the frame, as illustrated in figure \ref{fig:AGgeo}.  The volume around the detectors is filled with lead to shield ambient gamma rays and a thin sheet of copper lines the steel to protect the detectors from low-energy x-rays emitted from the lead.

Figure \ref{fig:AGgeo} also shows the positions of the detectors. If one views the target center as the origin of a coordinate system, the normal vector from the target surface points in the (1,-1,1) direction. The silicon detector is housed in a brass aperture case with a precision 27.6\,mm circular aluminum aperture, facing the deposit mounted in the target holder from the (1,-1,1) direction and about 80\,mm away. The HPGe detectors view the target from the top and bottom.

The Alpha-Gamma method requires accurate determination of neutron, alpha, and gamma losses, as well as consistency of the detection geometry throughout the measurement process. Thus, the target mounting structure was designed for accurate and repeatable positioning. The interchangeable target is held by a positioner attached to a flange that serves as a kinematic mount.  When the chamber is evacuated, air pressure on the exterior of the target positioning flange mates three pairs of parallel rods on the target positioning flange to tooling balls attached to the chamber.  The three parallel rod-sphere interactions restrict the target positioning flange to a unique spatial position.  The repositioning accuracy of the kinematic mount has been verified directly by measurement of the deposit center by theodolite and indirectly by activity measurements of alpha sources.

\subsection{Determination of the absolute neutron rate}

The calibration of the HPGe detectors begins with the determination of the absolute activity of an alpha-emitting source. ${}^{239}$Pu was selected as the isotope because of the simplicity of its energy spectrum, which is dominated by 5.2\,MeV alpha particles. Note that this energy is higher than that of the \Bt\ alpha particles, but this produces a negligible change in systematic effects such as Rutherford backscattering (see Section~\ref{sec:alphasys}). The alpha source is a ${5}$\,$\mu$g/cm${}^{2}$, 3\,mm diameter spot of PuO${}_{2}$ evaporated on an optically flat single crystal silicon wafer.  The alpha activity is determined by counting emitted alpha particles in a low solid-angle counting stack~\cite{Denecke99} of well-determined solid angle.  The low solid-angle counting stack consists of a spacer cylinder and a diamond-turned copper aperture.  The aperture diameter is measured by a coordinate measuring machine, and the distance between the source spot and the plane of the defining edge of the aperture is measured by a coordinate measuring microscope.  With these dimensions, the solid angle of the counting stack is precisely calculated.  The absolute alpha activity of the source $R_{\rm{Pu}}$ is therefore determined from the observed alpha-particle rate $r_{\rm{Pu,stack}}$ and the solid angle $\Omega_{\rm{stack}}$ where

\begin{equation}
R_{\rm{Pu}} = \frac {r_{\rm{Pu,stack}}}  {\Omega_{\rm{stack}}} .
\end{equation}

With the activity determined, the source is placed in the Alpha-Gamma device target position to establish the solid angle presented to the target center by the alpha detector.  The observed count rate $r_{\rm{Pu,AG}}$ and the known absolute activity $R_{\rm{Pu}}$ are used to determine the solid angle of the Alpha-Gamma silicon detector $\Omega_{\rm{AG}}$ where

\begin{equation}
\Omega_{\rm{AG}} = \frac {r_{\rm{Pu,AG}}}  {R_{\rm{Pu}}} .
\end{equation}

\noindent The source is replaced with a thin (${25}$\,$\mu$g/cm${}^{2}$) deposit of enriched, elemental \Bt\ prepared on a single crystal silicon substrate, and a neutron beam of total rate $R_{n}$ strikes the deposit.  The rate of neutron absorption $r_{\rm n, thin}$ in the deposit is given by

\begin{equation}
\label{eqn:rnthin}
r_{\rm n,thin} = R_{n}\sigma\rho_{N},
\end{equation}

\noindent where $\sigma$ is the \Bt\ absorption cross section and $\rho_{N}$ is the areal number density of the deposit.  Neutron absorption in \Bt\ produces ${}^{7}$Li and an alpha particle.  The ${}^{7}$Li nucleus is in an excited state 93.70\,\% of the time~\cite{Deruytter67,Stelts79} and will rapidly ($\tau = 73$ fs) de-excite by emission of a 478\,keV gamma ray.  This can be thought of as two separate reactions: an alpha-only reaction

\begin{equation}
\rm{n} + ^{10}\rm{B} \rightarrow ^{7}\rm{Li} (1015\, \rm{keV}) + \alpha(1776\, \rm{keV})
\end{equation}

\noindent  and an alpha plus gamma reaction (branching ratio $b_{\alpha\gamma}$ = 93.70\%)

\begin{eqnarray*}
\rm{n}+^{10}\rm{B}&\rightarrow ^{7}\rm{Li}^{*} + \alpha (1472\, \rm{keV})\\
                                      &\downarrow\\
                                      & ^{7}\rm{Li} (840\, \rm{keV}) + \gamma (478 \, \rm{keV}).
\end{eqnarray*}

The emitted alpha particles and gamma rays are detected in the silicon detector and HPGe detectors, respectively.  The observed rate of alpha particles $r_{\alpha, \rm{thin}}$ is

\begin{equation}
r_{\alpha,\rm{thin}} = \Omega_{\rm{AG}}r_{\rm n, thin},
\end{equation}

\noindent and the observed rate of gamma rays $r_{\gamma,\rm{thin}}$ is

\begin{equation}
r_{\gamma, \rm{thin}} = \epsilon_{\gamma} b_{\alpha\gamma} r_{\rm n, thin},
\end{equation}

\noindent where $\epsilon_{\gamma}$ is the detection efficiency for 478\,keV gamma rays.  The observed alpha rate and the known solid angle are used to find the neutron absorption rate

\begin{equation}
r_{n, \rm{thin}} = \frac{r_{\alpha,\rm{thin}}}{\Omega_{\rm{AG}}}.
\end{equation}

\noindent This is, in turn, used to determine the $\gamma$ detection efficiency

\begin{equation}
\label{eqn:epsilon}
\epsilon_{\gamma} = \frac{r_{\gamma,\rm{thin}}}{b_{\alpha\gamma}r_{\rm n, thin}} = \frac{1}{b_{\alpha\gamma}}\frac{r_{\gamma,\rm{thin}}}{r_{\alpha,\rm{thin}}}\Omega_{\rm{AG}}.
\end{equation}

\noindent  At this point, the thin \Bt\ deposit is replaced with a thick target of 98\,\% enriched ${}^{10}$B${}_{4}$C, and one is ready to measure the neutron beam rate.  The entire beam is absorbed in the target, and the measured gamma rate is

\begin{equation}
\label{eqn:thickgammarate}
r_{\gamma,\rm{thick}} = \epsilon_{\gamma}b_{\alpha\gamma}R_{n}.
\end{equation}

\noindent  Using equation \ref{eqn:epsilon}, one can express $R_{n}$ as

\begin{equation}
\label{eqn:Rn}
R_{n} = \frac{r_{\gamma,\rm{thick}}}{\epsilon_{\gamma}b_{\alpha\gamma}} = r_{\gamma,\rm{thick}}\frac{r_{\alpha,\rm{thin}}}{r_{\gamma,\rm{thin}}}\frac{1}{\Omega_{\rm{AG}}}.
\end{equation}

\noindent The crux of this result is that one determines the neutron rate entirely in terms of measured quantities, without reference to $b_{\alpha\gamma}$ or other input parameters.

\section{Neutron flux monitor}\label{sec:FM}
\subsection{Principle of operation}

An important application of the black detector was to measure directly the detection efficiency of a ${}^{6}$Li-based neutron flux monitor~\cite{Williams89} and to reduce significantly the uncertainty in that quantity. We describe the measurement principle and construction of the apparatus here. The monitor measures the capture flux of a neutron beam by counting the alpha or triton emitted in neutron absorption on a thin ${}^{6}$LiF deposit via the reaction 

\begin{equation*}
\rm{n} + {}^{6}\rm{Li} \rightarrow \alpha (2070\, \rm{keV}) + {}^{3}\rm{H} (2720\, \rm{keV}).
\end{equation*}

\noindent The existing technique for determining the detection efficiency of the monitor was limited to a precision of 0.3\,\% by uncertainty in the \Ls\ cross section and the mass of the lithium deposit~\cite{Nico05}. While this precision is acceptable for the majority of uses for the flux monitor, it is not sufficient for the sub-0.1\,\% applications described herein. One notes that the \Ls\ cross section is not determined by the user of the monitor, but it is an input that is obtained from a database of evaluated neutron cross sections, and therefore, one's determination of a given neutron flux is a function of the value of the evaluated cross section. This situation is unacceptable for high precision work. Because the flux monitor uses a thin deposit, one can place both the flux monitor and the Alpha-Gamma device simultaneously on a monochromatic neutron beam to determine precisely the efficiency of the flux monitor without reference to either deposit mass or the \Ls\ cross section~\cite{Yue2011}. 

\begin{figure}[htbp]
  \centering
  \includegraphics[width=1\textwidth]{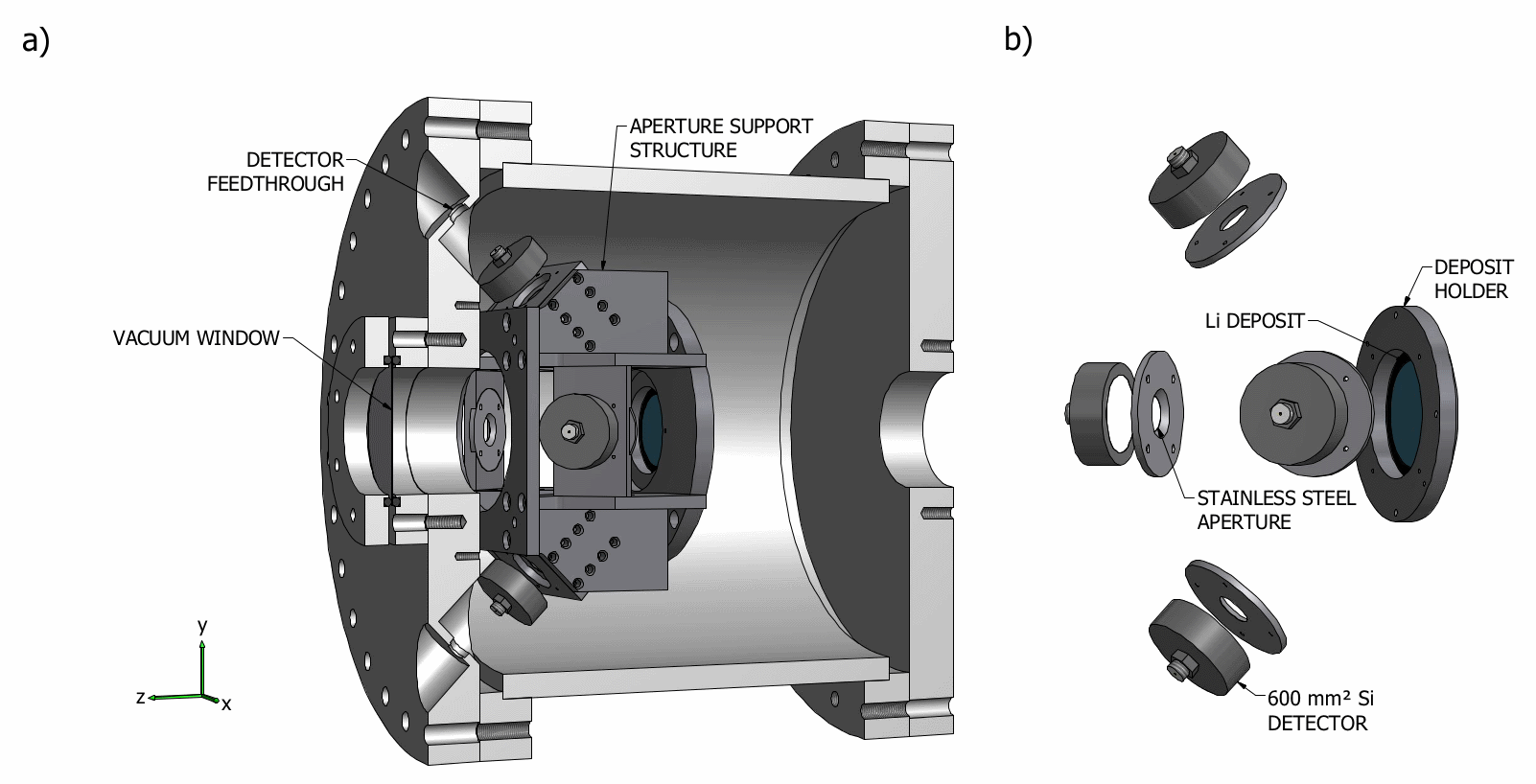}
  \caption{Illustrations of a) a section view of the neutron monitor vacuum chamber and b) a representation of the detection geometry.}
  \label{fig:FMDetGeo}
\end{figure}

The neutron flux monitor is illustrated schematically in figure \ref{fig:FMDetGeo}. There are two nearly-identical monitors in existence; they are interchangeable so that one device could be used for a neutron lifetime measurement while the other device underwent calibration. A rigid frame holds a thin ${}^{6}$LiF deposit fixed with respect to four precision ground apertures. The apertures mask four charged-particle detectors and define the solid angle for detection of the reaction alphas and tritons.  The detectors are silicon charged-particle detectors positioned in the four cardinal directions, and each faces the target deposit at an angle of 45${}^{\circ}$.  The neutron monitor is characterized by an efficiency parameter $\epsilon(v)$ that denotes the ratio of detected reaction products to incident neutrons of velocity $v$

\begin{equation}
\epsilon(v) = \frac{2N_{A}}{4\pi{A}}\sigma(v)\int\int{\Omega_{\rm{FM}}(x,y)\rho(x,y)\phi(x,y)\rm{d}x\rm{d}y},
\end{equation}

\noindent where $N_{A}$ is the Avogadro constant, $A = 6.01512$ g/mol is the atomic weight of ${}^{6}$Li, $\sigma(v)$ is the ${}^{6}$Li(n,t)${}^{4}$He cross section for a neutron of velocity $v$, $\Omega_{\rm{FM}}(x,y)$ is the monitor solid angle, $\rho(x,y)$ is the areal density of the ${}^{6}$Li in the deposit, and $\phi(x,y)$ is the areal distribution of the neutron intensity incident on the target. The coordinates $x$ and $y$ are on the face of the deposit, perpendicular to the beam direction. The neutron monitors may be operated on both monochromatic and polychromatic neutron beams of many different beam sizes, and so it is judicious to characterize the efficiency of the detector for a particular configuration.  We define $\epsilon_{0}(0,0)$ to be the detection efficiency of the monitor for a beam of monochromatic thermal neutrons ($v_{0} = 2200$\,m/s) infinitely narrow in extent and striking the center of the deposit ($\phi(x,y) = \delta(x)\delta(y)$) such that

\begin{equation}
\label{eqn:e0calc}
\epsilon_{0}(0,0) = \frac{2N_{A}}{A}\sigma_{0}{\Omega_{\rm{FM}}(0,0)\rho(0,0)},
\end{equation}

\noindent where $\sigma_{0}$ is the ${}^{6}$Li thermal neutron cross section.  From $\epsilon_{0}(0,0)$, the efficiency of the monitor for any beam may in principle be determined. The solid angle subtended by the apertures was measured mechanically by a coordinate measuring machine and experimentally by using a calibrated $\alpha$-source designed to fit in the deposit holder. The two measurement techniques agreed to better than the measurement uncertainty of 0.03\,\%, and the value of $\Omega_{\rm{FM}}(0,0) = (4.2021 \pm 0.0014) \times{10}^{-3}$ (in units of $4\pi$) was used for the solid angle of the monitor characterized in this work. 

In this direct method for determining the neutron monitor efficiency, the monitor and Alpha-Gamma device are operated on a beam of wavelength $\lambda_{\rm{mono}}$ and total rate $R_{\rm n}$.  The observed rate of alphas and tritons ($r_{\alpha,t}$) is

\begin{equation}
r_{\alpha,t} = \epsilon{R_{\rm n}},
\end{equation}

\noindent where $\epsilon$ is the detection efficiency of the monitor for neutrons of wavelength $\lambda_{\rm{mono}}$.  The total neutron rate can be obtained from $r_{\gamma,\rm{thick}}$ and observables from the Alpha-Gamma calibration procedure (equation \ref{eqn:Rn}), and $\epsilon$ is then determined

\begin{equation}
\epsilon = \frac{r_{\alpha,t}}{r_{\gamma, \rm{thick}}}\frac{r_{\gamma, \rm{thin}}}{r_{\alpha,\rm{thin}}}\frac{r_{\rm{Pu,\rm{AG}}}}{R_{\rm{Pu}}}.
\end{equation}

\noindent Under the assumption of a $1/v$ cross section, a measurement of $\lambda_{\rm{mono}}$ then allows one to determine the detection efficiency for an equivalent beam of thermal neutrons of wavelength $\lambda_{0}$ (corresponding to $v_{0} = 2200$ m/s)
\begin{equation}
\label{eqn:epsiloneq}
\epsilon_{0} = \epsilon\frac{\lambda_{0}}{\lambda_{\rm{mono}}} = \frac{r_{\alpha,t}}{r_{\gamma,\rm{thick}}}\frac{r_{\gamma,\rm{thin}}}{r_{\alpha,\rm{thin}}}\frac{r_{\rm{Pu,\rm{AG}}}}{R_{\rm{Pu}}}\frac{\lambda_{0}}{\lambda_{\rm{mono}}}.
\end{equation}
With knowledge of $\phi\left(x,y\right)\rho\left(x,y\right)$, the idealized efficiency $\epsilon_{0}(0,0)$ can be calculated.

\subsection{${}^{6}$LiF deposits}

Careful fabrication and characterization of the ${}^{6}$LiF deposits for the neutron monitor were a critical component of the flux monitor development. They were produced and characterized in a joint effort between the National Institute of Standards and Technology (NIST) and the Institute for Reference Materials and Measurements (IRMM) in Geel, Belgium.  The ${}^{6}$Li layer was deposited onto silicon wafers with a custom evaporation rig based on a rotating multi-substrate holder~\cite{Pauwels95}.  A tantalum crucible filled with ${}^{6}$LiF was placed approximately 40 cm from the rotator and heated to evaporative temperature. Seven substrates (six silicon and one stainless steel) were held at normal incidence to the evaporative particle flux.  The entire substrate holder orbited the crucible (``yearly'' rotation), and the individual holders rotated about the axis established by the evaporator and substrate holder (``daily'' rotation).  The two rotation periods were chosen to minimize the effect of asymmetry in the evaporated particle flux on the uniformity of the deposit.  Each substrate holder was masked with a precision aperture.   To ensure that the apertures were flush to the surface of the substrate, each was prepared with optical grinding methods.  The diameter of each deposit was controlled by ensuring masking aperture bore uniformity.  The final bore enlargement was performed by clamping pairs of apertures together and grinding to the desired diameter of 38\,mm.  

The areal density of the deposit was determined by measuring the amount of ${}^{6}$Li present in the deposit and the shape of the deposit.  The deposit profile was measured by a visible light spectrophotometer and was calculated from the known dimensions of the evaporation rig and the rotation speeds.  The measured profile verified the derived profile.  The sharpness of the deposit edge was measured by microscope and Talistep recording, and the deposit diameter was measured by an Abbe-comparator~\cite{Pauwels95}.

The ${}^{6}$LiF deposits were prepared in three evaporations of nominal areal densities of 20\,$\mu$g/cm${}^{2}$, 30\,$\mu$g/cm${}^{2}$, and 40\,$\mu$g/cm${}^{2}$. We note that most numerical values presented in this paper are for the 40\,$\mu$g/cm${}^{2}$ deposit, but they are very representative of the lighter deposits. (The heaviest deposit was selected because it was used in the neutron lifetime experiment~\cite{Dewey2003}). A combination of relative reaction rate comparison and isotope dilution mass spectrometry (IDMS) was used to determine the amount of ${}^{6}$Li in the deposits~\cite{Scott95}.  The alpha and triton reaction rate for each deposit was measured on a thermal neutron beam using a device similar to the NIST neutron flux monitor.  The reaction rates were used to establish the relative mass difference between the deposits of the same nominal mass.  Two deposits from each evaporation were then destructively analyzed by IDMS for absolute measurement of their masses.  The reaction rate data and absolute mass determination from the sacrificed deposits established the reaction rate per unit mass.  The absolute mass of each of the remaining deposits was then determined from their measured reaction rates. The ${}^{6}$LiF deposit most commonly used in the neutron flux monitor was determined to have an average areal density $\bar{\rho} = 39.3$ $\mu\rm{g/cm}^{2} \pm 0.25$\,\%.  The deposit profile as a function of radial position (in mm) from the deposit center is

\begin{equation}
\label{eqn:rhor}
\rho(r) = \bar{\rho}\frac{1-(1-0.995)\left(\frac{r}{19}\right)^{2}}{1-\frac{0.005}{2}},
\end{equation}

\noindent and thus the areal density at the center of the deposit is $\rho(0,0) = 39.40$ $\mu\rm{g/cm}^{2} \pm 0.25$\,\%.

As noted, the ${}^{6}$Li thermal neutron cross section must be taken from evaluated nuclear data files (ENDF).  The most recent evaluation is ENDF/B-VII, which reports $\sigma_{0} = (938.5  \pm 1.3$)\,b~\cite{ENDF7}.  This 0.14\,\% uncertainty comes largely from the combined-analysis uncertainty from $R$-matrix evaluations.  The ENDF-determined ${}^{6}$Li(n,t) thermal neutron cross section used does not come from one precision measurement at thermal neutron energy but instead from a global evaluation of many neutron reactions, mostly at much higher energy. The two most recent evaluations of the ${}^{6}$Li cross section are limited to an uncertainty of 0.14\,\% and are in slight disagreement with one another. The cross section is the only quantity that goes into the determination of the monitor efficiency that does not come from a first-principles measurement. Using $\sigma_{0} = (938.5 \pm 1.3)$\,b, $\Omega(0,0) = (4.2021 \pm 0.0014) \times{10}^{-3}$, $\rho(0,0) = (39.40 \pm 0.10)$\,$\mu$g/cm${}^{2}$, and equation \ref{eqn:e0calc}, we find $\epsilon_{0}(0,0) = (3.1111  \pm 0.0089) \times{10}^{-5}$.  This 0.29\,\% uncertainty is likely near the limit of the techniques used for this measurement. The ${}^{6}$LiF deposit areal density determination (0.25\,\%) is the result of an extensive measurement campaign. 

The inherent limitations of the cross section and areal density measurements underscore the desire to develop a method of determining the flux monitor efficiency that does not depend upon either of these quantities. By using the Alpha-Gamma device to measure the neutron rate simultaneously with  the neutron monitor, one establishes the efficiency of the neutron monitor independent of the \Ls\ cross section, the deposit areal density, and the solid angle of the particle detectors. Instead, the method utilizes absolute particle counting techniques and requires a precise knowledge of the neutron wavelength, as seen in Eqn.~\ref{eqn:epsiloneq}.

\section{Data acquisition and analysis}
\label{sec:DAQ}

\subsection{Data runs}

The measurements presented in this work were performed at the NIST Center for Neutron Research (NCNR). The NCNR operates the NBSR, a 20\,MW, D${}_{2}$O-moderated research reactor that provides thermal neutrons to nine experimental stations and cold neutrons by moderation in liquid hydrogen~\cite{Rush2011}. Neutron guide NG-6 was operated by the Physical Measurement Laboratory at NIST for the study of fundamental neutron physics and neutron dosimetry~\cite{NicoJRES05}.  In addition to the polychromatic end station beam, three monochromatic neutron beams are generated upstream of NG-6 by Bragg reflection from appropriate monochromator crystals. Their wavelengths are nominally 0.89\,nm, 0.496\,nm, and 0.383\,nm for beamlines NG-6u, NG-6m, and NG-6a, respectively. The neutron flux and wavelength measurements were carried out on NG-6m. The neutron flux of the beam varied depending on specific experimental conditions but was typically a few times $10^5$/s. 

\begin{figure}
  \centering
  \includegraphics[width=1\textwidth]{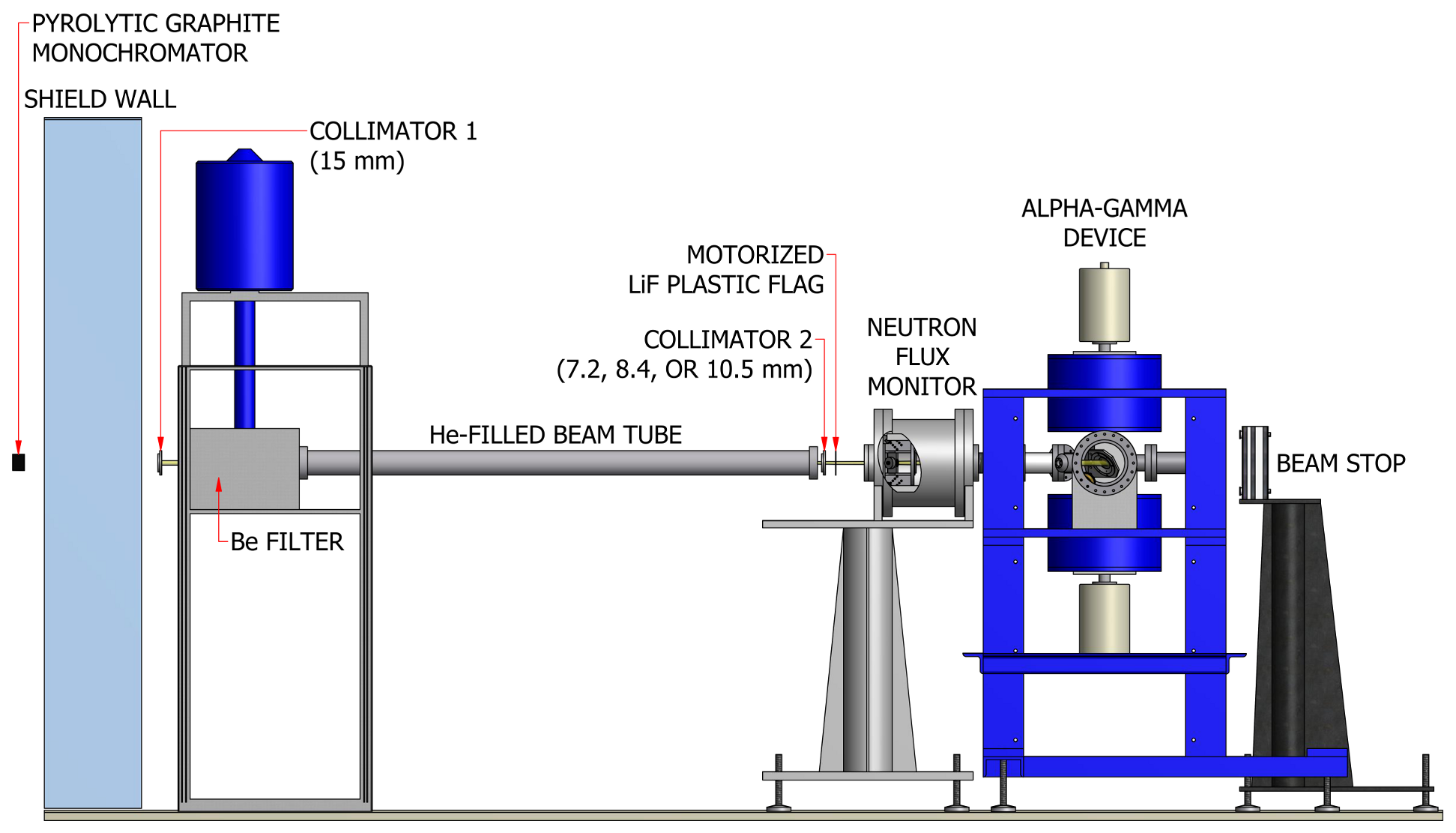}
  \caption{Illustration of the experimental setup on the NG-6m beamline. Cold neutrons enter from the left and pass through the filter where they are collimated in a He-filled flight tube. After exiting the flight tube, neutrons pass through the flux monitor and Alpha-Gamma device. The gap between the flight tube and flux monitor allows for the insertion of components to measure the wavelength of the beam.}
  \label{fig:NG6mlayout}
\end{figure}

The  layout of Alpha-Gamma device and the flux monitor is shown in figure \ref{fig:NG6mlayout}.  A pyrolytic graphite monochromator was used to diffract the 0.496\,nm neutrons used for NG-6m from the polychromatic beam NG-6. The beam passed through a 15\,mm diameter sintered ${}^{6}$LiF ceramic collimator before entering a polycrystalline beryllium filter.  The filter preferentially scatters neutrons of wavelength below 0.396\,nm, effectively removing $\lambda/2$, $\lambda/3$, and higher order Bragg reflections from the beam.  A helium-filled guide tube efficiently transported the neutrons to a second ${}^{6}$LiF collimator whose diameter was varied (7.2\,mm, 8.4\,mm, or 10.5\,mm) for the study of systematic effects.  A motorized ${}^{6}$LiF-plastic flag was used for beam modulation to obtain periodic background measurements.  A 10\,cm air gap was left between the final collimator and the entrance to the flux monitor to accommodate the neutron wavelength measuring components, critical for determining the flux monitor efficiency.

Data were acquired for the neutron monitor efficiency from June 2010 to December 2010. Each data set consisted of a 3-day cycle of measurements. The statistical uncertainty was optimized when each measurement was performed for one day.  On the first day, a thin-target measurement establishes the initial alpha/gamma ratio.  On the second day, the thick target measurement determines the absolute flux of neutrons.  On the third day, the final alpha/gamma ratio is measured using the thin target.  The purpose of performing two thin-target measurements is to eliminate first-order drifts in gamma detector efficiency.

\subsection{Data acquisition system}

The experiment requires accurate counting of a broad range of charged particles and a narrow energy region of gamma rays. The data acquisition system (DAQ) should perform straightforward particle counting with a minimum number of potential complexities that might confound the determination of any given rate. Deadtime corrections must also be straightforward to calculate. To accomplish this goal, the data acquisition system was designed to minimize complexity.

\begin{figure}[htbp]
  \centering
  \includegraphics[width=1\textwidth]{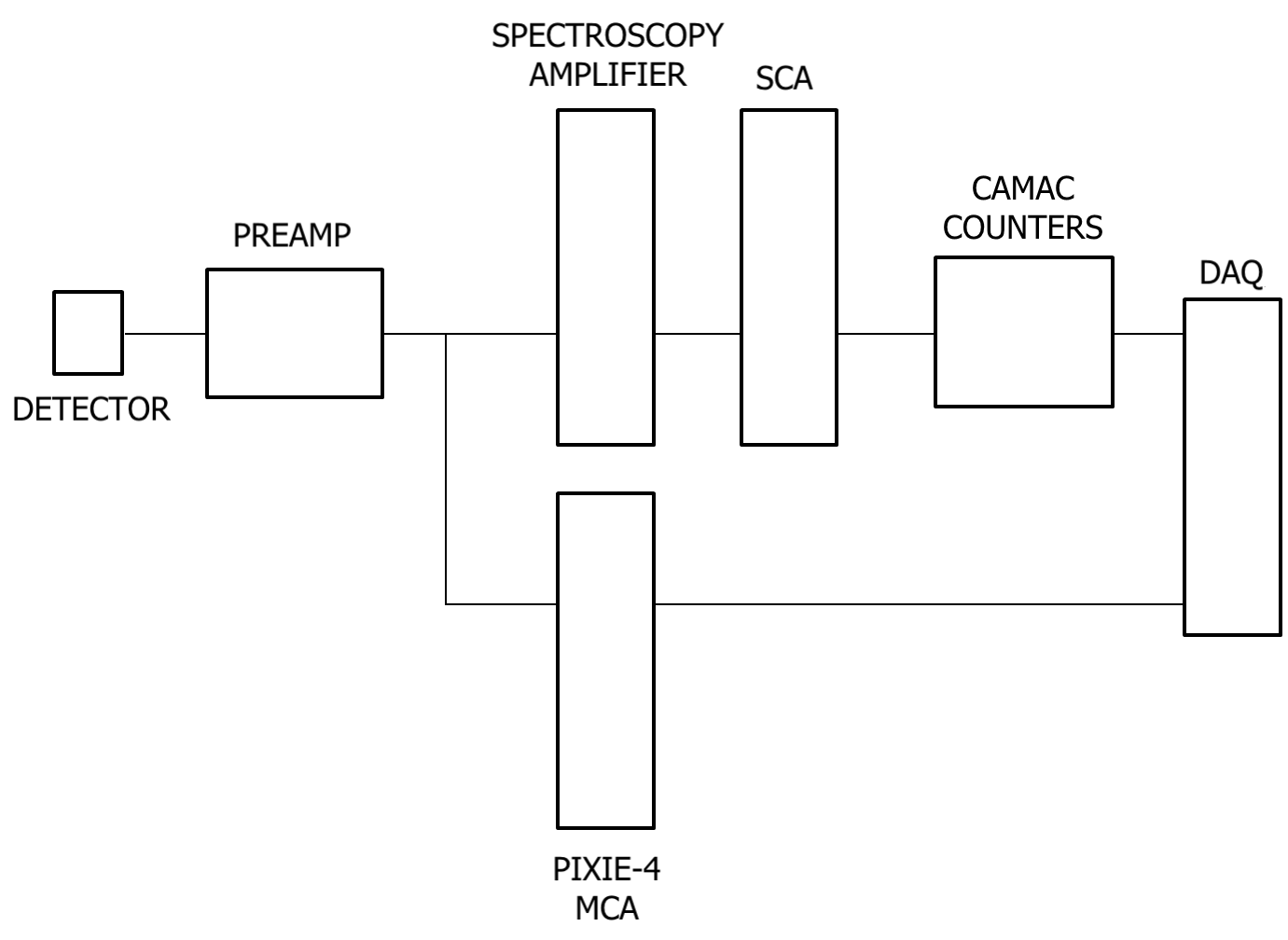}
  \caption{Block diagram of a detector's counting and spectroscopy electronics.}
  \label{fig:WiringDiagram}
\end{figure}

The charged particle counting in the Alpha-Gamma device was performed with a single 900\,mm${}^{2}$ silicon charged-particle detector while the neutron flux monitor required four 600\,mm${}^{2}$ silicon charged-particle detectors.  The gamma rays were detected in two 20\,\% relative efficiency HPGe detectors.  The counting electronics were analog NIM and CAMAC electronics, preferable for their speed and simplicity. For both the particle and gamma signal, the impulse from the detector went into a preamplifier whose output was split, one output going directly to an XIA Pixie-4 multichannel analyzer (MCA) and the other going to a spectroscopy amplifier.  The amplifier signal was read into single channel analyzers (SCA) for pulse-height discrimination.  Pulses meeting the necessary thresholds sent TTL pulses from the SCA to the appropriate channel in a CAMAC hex counter, as shown in figure~\ref{fig:WiringDiagram}.

The typically 20\,mV to 100\,mV preamplifier tail pulses are converted to about 1\,V Gaussian pulses by a spectroscopy amplifier.  An SCA operating in normal mode has two independent thresholds (lower and upper level).  The Gaussian pulse is read in and the SCA puts out a TTL pulse on the lower or upper level output as the signal rises over the respective threshold.  The threshold does not become live again until the signal falls below the threshold value.  Peak summing is accomplished by setting the lower and upper thresholds around a signal peak.  The peak sum is given by the difference between the lower and upper sums.

\begin{figure}[htbp]
  \centering
  \includegraphics[width=1\textwidth]{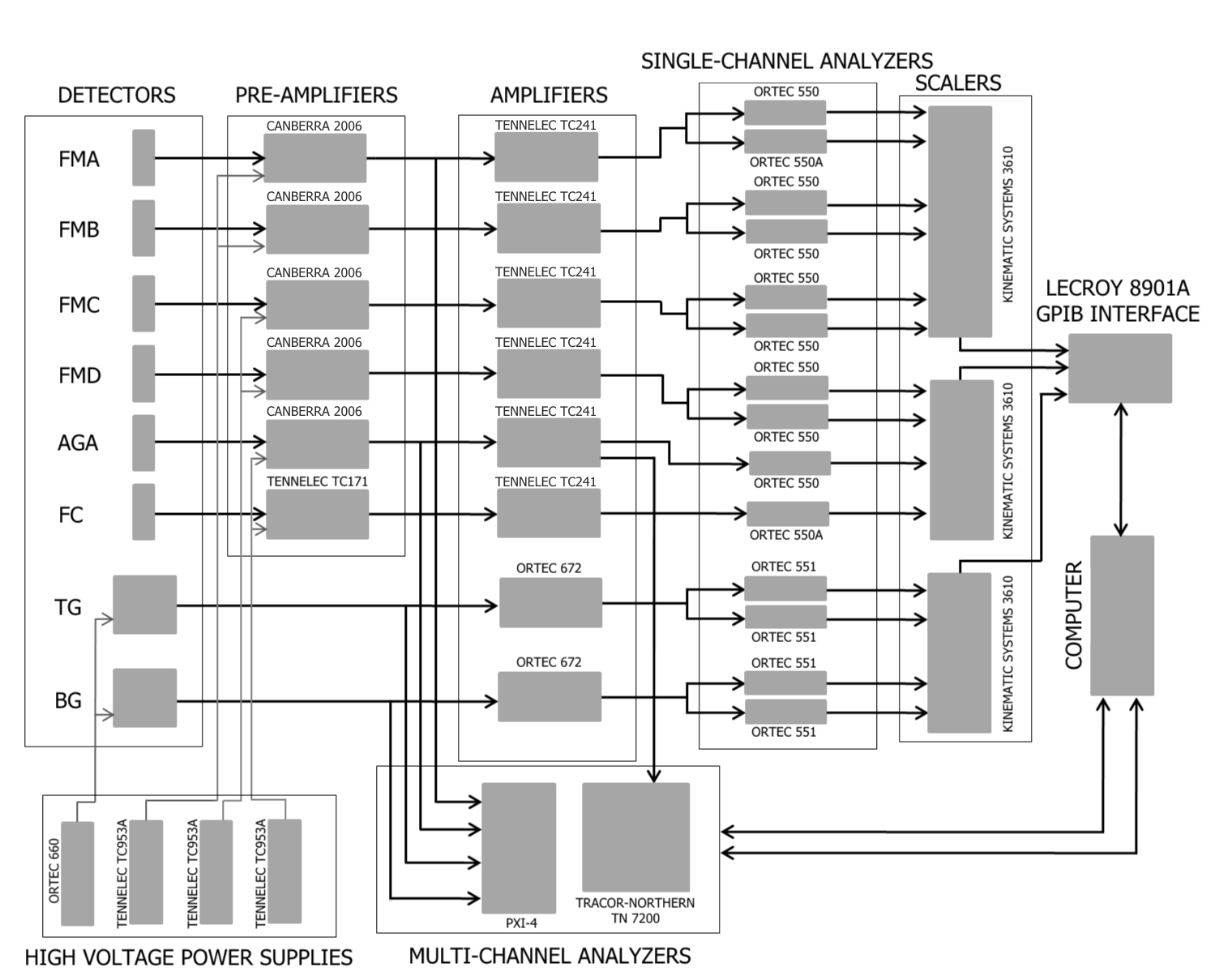}
  \caption{Electronics diagram for the Alpha-Gamma device and neutron monitor.  FMA - FMD are the four neutron monitor silicon detectors, AGA is the Alpha-Gamma silicon detector, FC is an upstream fission chamber and TG and BG are the top and bottom HPGe detectors.}
  \label{fig:AGdevelec}
\end{figure}

A block diagram of the apparatus electronics is shown in figure \ref{fig:AGdevelec}.  The DAQ software code was written in LabWindows/CVI, and its primary function was communicating with a CAMAC crate via a GPIB controller. The program reads out 15 hex scalers every minute of the computer clock.  The time between readouts can vary due to processor load, so a CAMAC millisecond timer is tracked with a hex scaler counter.  The DAQ also communicates with a digital multimeter via GPIB to monitor either the temperature of the beryllium filter or the bias shutdown signal on one of the gamma detectors.  An Input Gate/Output Register module in the CAMAC crate is used to control the modulation state of the upstream lithium flag, monitor status of the local beam shutter, and monitor the gamma detector liquid nitrogen fill system.

In addition to the scaler counting, four signals (the two gamma detectors, the Alpha-Gamma silicon detector, and channel A of the neutron monitor) were read in by a Pixie-4 module, a digital waveform acquisition card.  Each channel is digitized by a 14-bit 75\,MHz analog-to-digital converter (ADC).  The Pixie-4 is operated by the LabWindows/CVI data acquisition program.  Each minute spectra with 16k channels are acquired and buffered into memory.  At the end of every shutter cycle (typically 15 minutes of beam on data followed by 5 minutes of beam off), a beam on and beam off spectrum are written to a file.  The DAQ keeps an accumulated beam-on and beam-off spectrum on display for quick diagnostics.

\subsection{Data analysis}

Recall from Section~\ref{sec:prinofmeasure} that there is a sequence of four procedures that are required to determine the efficiency of neutron counting from the Alpha-Gamma device. Initially, the efficiency of its alpha detector is established using the well-characterized ${}^{239}$Pu alpha source. The source is then removed and replaced with a thin \Bt\ deposit. The beam shutter is opened and neutrons are incident on the target, and the well-determined efficiency of the alpha detector is transferred to the gamma-ray detectors. Finally, the thick \Bt\ target is placed in the beam to measure the total neutron flux by counting the gamma rays. The fundamental goal of the data analysis consists of converting these charged particle and gamma-ray energy spectra from each of these steps into total count rates. A good understanding of the background spectra is critical to obtaining the correct rates.

\begin{figure}
  \centering
  \includegraphics[width=\textwidth]{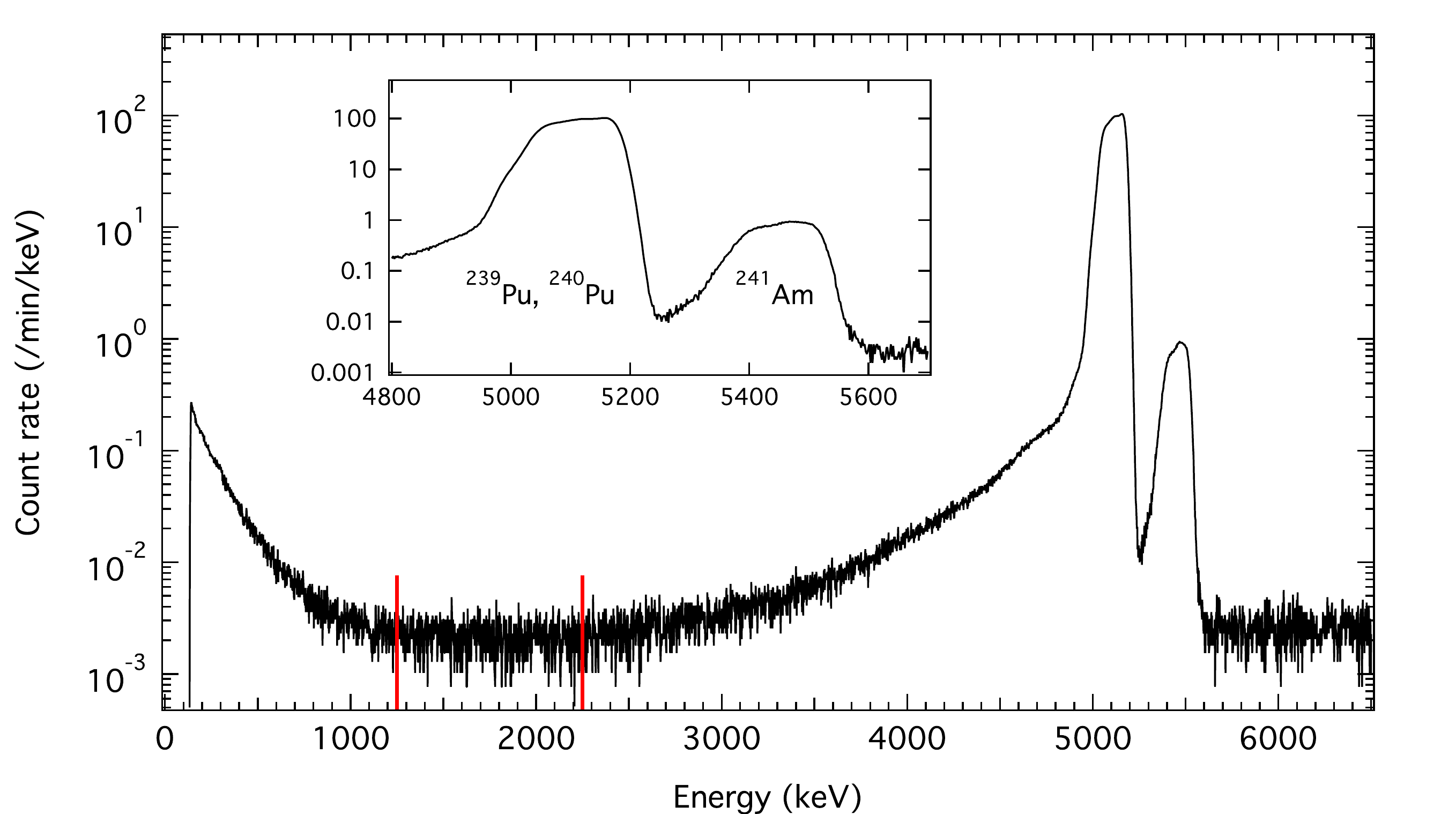}
  \includegraphics[width=\textwidth]{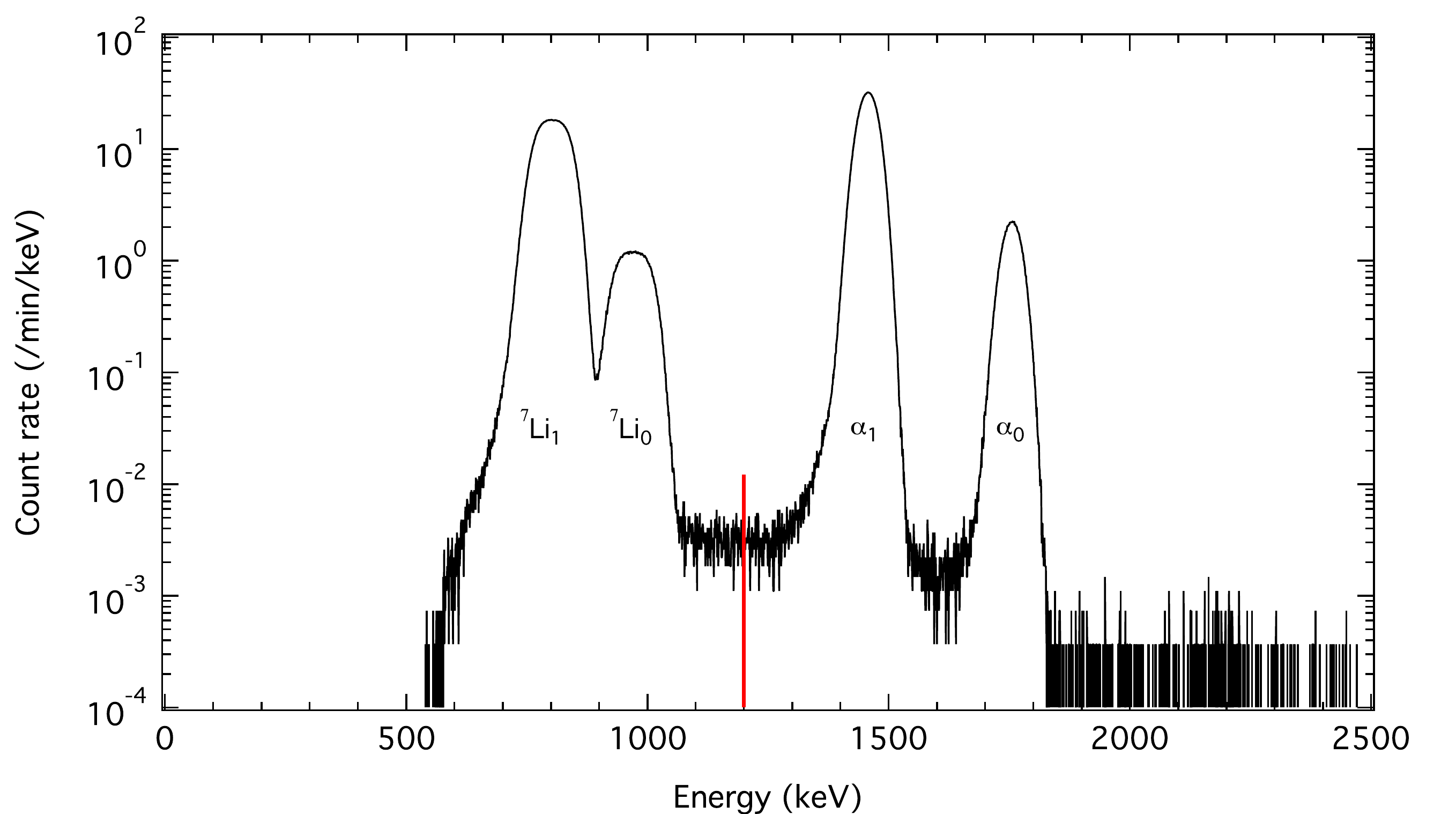}
  \caption{Top: charged particle spectrum from the ${}^{239}$Pu source. The inset indicates other isotopes that contribute to the alpha rate. The red lines indicate a typical region used for background subtraction. Bottom: charged particle spectrum from the thin \Bt\ target. The red line indicates the position of a typical analysis threshold for counting the alpha particles.}
  \label{fig:Pualphaspec}
\end{figure}

For the alpha counting, the ${}^{239}$Pu alpha source was mounted in the target holder of the Alpha-Gamma device (see figure~\ref{fig:AGgeo}). The surface area of the detector illuminated by the alpha particles was assumed to have unit detection efficiency.  A typical energy spectrum from the source is shown in figure~\ref{fig:Pualphaspec}. In addition to the ${}^{239}$Pu alpha particles, there are also contributions from contaminants of ${}^{240}$Pu and ${}^{241}$Am. The fraction of contaminants is not large and does not effect the determination of the efficiency because the relevant parameter is total number of alpha particles emitted from the source. The detector background and noise are very small contributions that were subtracted from the spectrum in a straightforward manner and did not require frequent measurement.

After completion of the source measurement, the alpha source is removed and the thin \Bt\ deposit is placed in the target holder. When the beam shutter is opened, the same silicon detector and aperture measures the  ${}^{10}$B(n,$\alpha$)${}^{7}$Li charged particle spectrum. Figure \ref{fig:Pualphaspec} shows the resulting charged-particle spectrum and illustrates the high signal-to-background ratio of the two alpha peaks and the good resolution from the detector noise. The two recoil peaks of the ${}^{7}$Li  are also clearly resolved. In principle, the ${}^{7}$Li peaks could be counted as well, but they fall on the noise tail. Additionally, the $\alpha$ signal has the highest rate of all the signals in the thin target mode, so increasing its rate by accepting an additional peak has little statistical benefit.

\begin{figure}
  \centering.pdf
  \includegraphics[width=\textwidth]{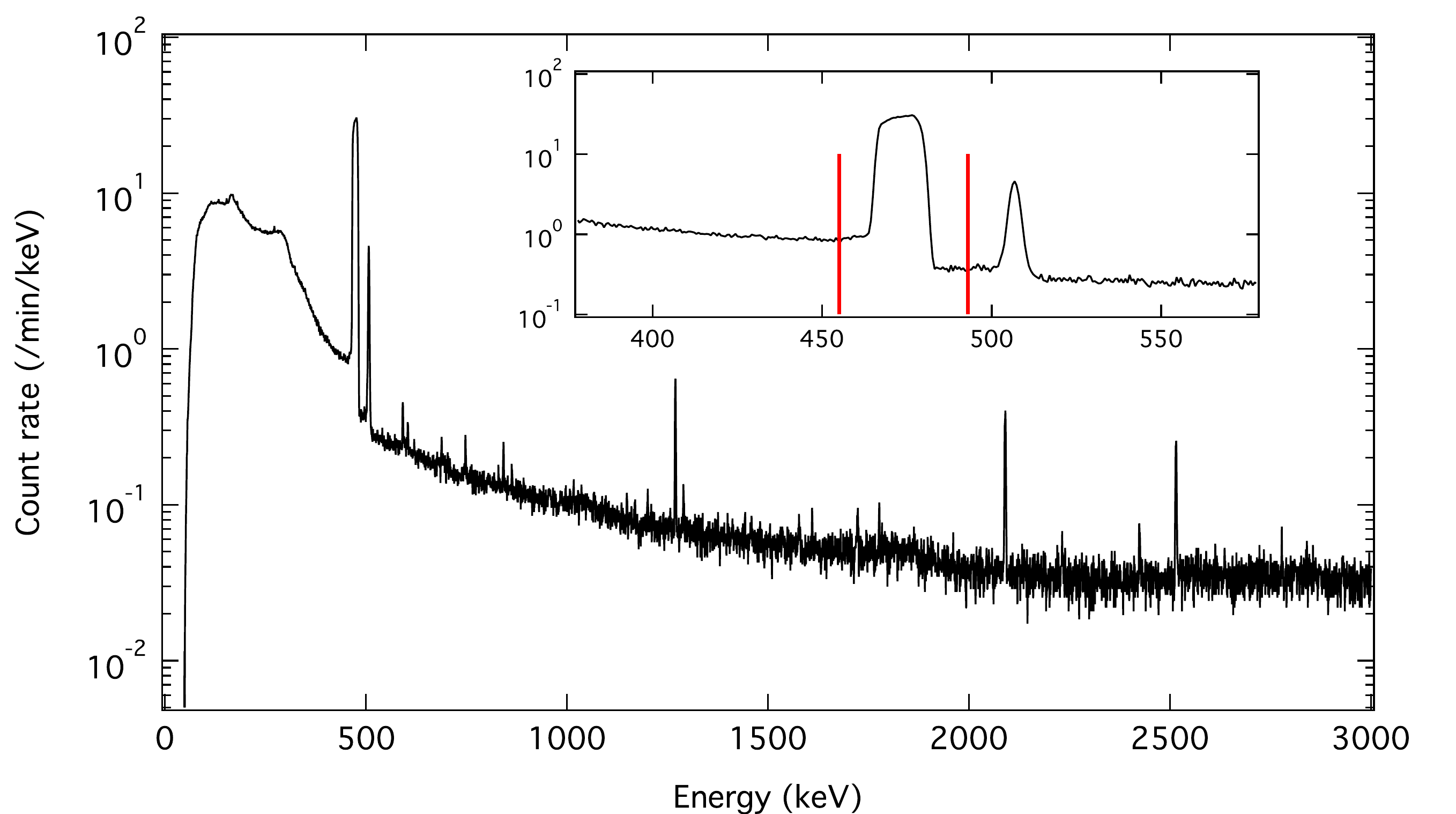}
  \includegraphics[width=\textwidth]{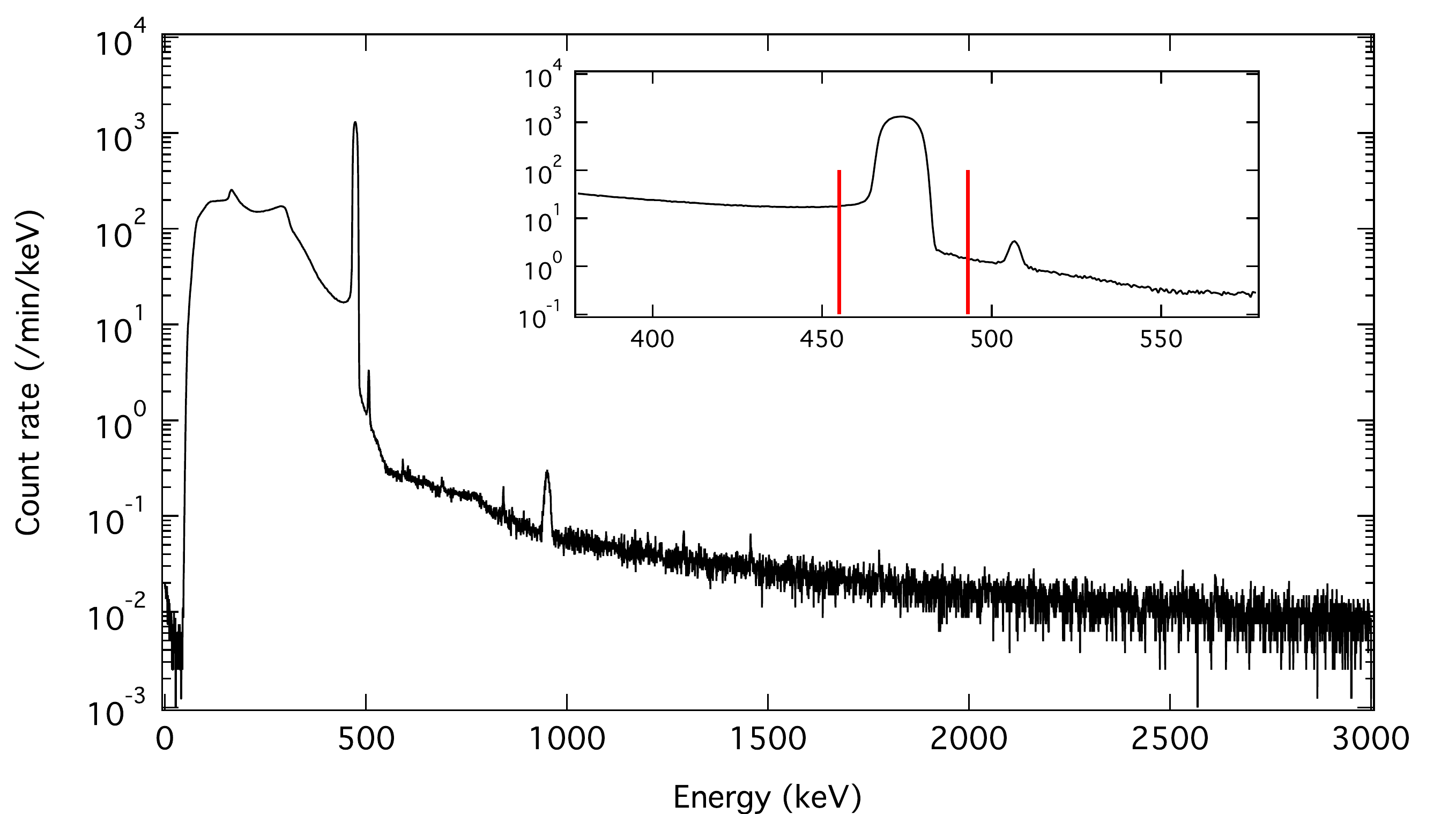}
  \caption{Top: plot of the gamma-ray spectrum from the thin \Bt\ target. The inset shows the region expanded around the 478\,keV line, and the red lines indicate the analysis region. Bottom: plot of the gamma-ray spectrum from the thick \Bt\ target. The inset shows the region expanded around the 478\,keV line, and the red lines indicate a typical analysis region.}
  \label{fig:gammaspec}
\end{figure}

Simultaneously, one acquires the gamma-ray spectrum from the same reaction to permit the transfer of the alpha detection efficiency to the gamma-ray detectors. A typical thin \Bt\ target gamma-ray spectrum is shown in figure \ref{fig:gammaspec}.  Prompt gamma lines from Si activation can be seen, as well as small amounts of background gammas from ambient environmental radioactivity (e.g., Ra, K, etc.).  The 478\,keV gamma from capture on \Bt\ is broadened by the distribution of Doppler shifts caused by the relativistic energy of the ejected ${}^{7}$Li nucleus.  The signal peak is still clearly resolved from the electron-positron annihilation peak present at 511\,keV and is only an order of magnitude resolved from background gammas.  The thick \Bt\ target gamma spectrum is shown in figure \ref{fig:gammaspec}.  The rate in the 478\,keV peak is significantly higher, and thus the line is better resolved from the background.

The gamma-ray background comprises a significant fraction of the gamma signal in the thin target during operating mode.  Because the gamma counting is essential for both \Bt\ targets, all of the data (aside from the $^{239}$Pu source measurements) include frequent, dedicated measurements of the beam-off background.  The motorized ${}^{6}$Li-loaded plastic flag is used to modulate the beam in a 15-minute-on, 5-minute-off cycle. If $N_{\rm{on}}$ and $N_{\rm{off}}$ are the observed number of counts for a signal for the beam on and beam off durations of pulser-determined time $T_{\rm{on}}$ and $T_{\rm{off}}$, the background subtracted average signal rate $r$ for the data cycle is

\begin{equation}
r = \frac{N_{\rm{on}}}{T_{\rm{on}}} - \frac{N_{\rm{off}}}{T_{\rm{off}}}.
\end{equation}

\noindent The data cycle average rate for each signal is recorded and a run average rate is determined.  The rate from the top and bottom gamma detectors ($r_{\rm{T}\gamma}$ and $r_{\rm{B}\gamma}$) are combined to form a geometric mean

\begin{equation}
r_{\gamma} = \sqrt{{r_{\rm{T}\gamma}}{r_{\rm{B}\gamma}}}.
\end{equation}

\noindent Using equation \ref{eqn:epsiloneq}, we find that the beam-related statistical accumulation is reduced to the measurement of two ratios: $\frac{r_{\gamma,\rm{thin}}}{r_{\alpha,\rm{thin}}}$ when the thin \Bt\ target is in the Alpha-Gamma device and $\frac{r_{\alpha,t}}{r_{\gamma,\rm{thick}}}$ when the thick \Bt\ target is in use.  The statistical accumulation for each $\epsilon$ measurement is composed of three one-day measurements in a thin target-thick target-thin target pattern in the Alpha-Gamma device.  This provides two measurements of $\epsilon_{\gamma}$ per flux measurement with the thick target, eliminating the effect of first-order drifts in gamma detection efficiency.

\begin{figure}[h]
	\centering
	\includegraphics[width=1.0\textwidth]{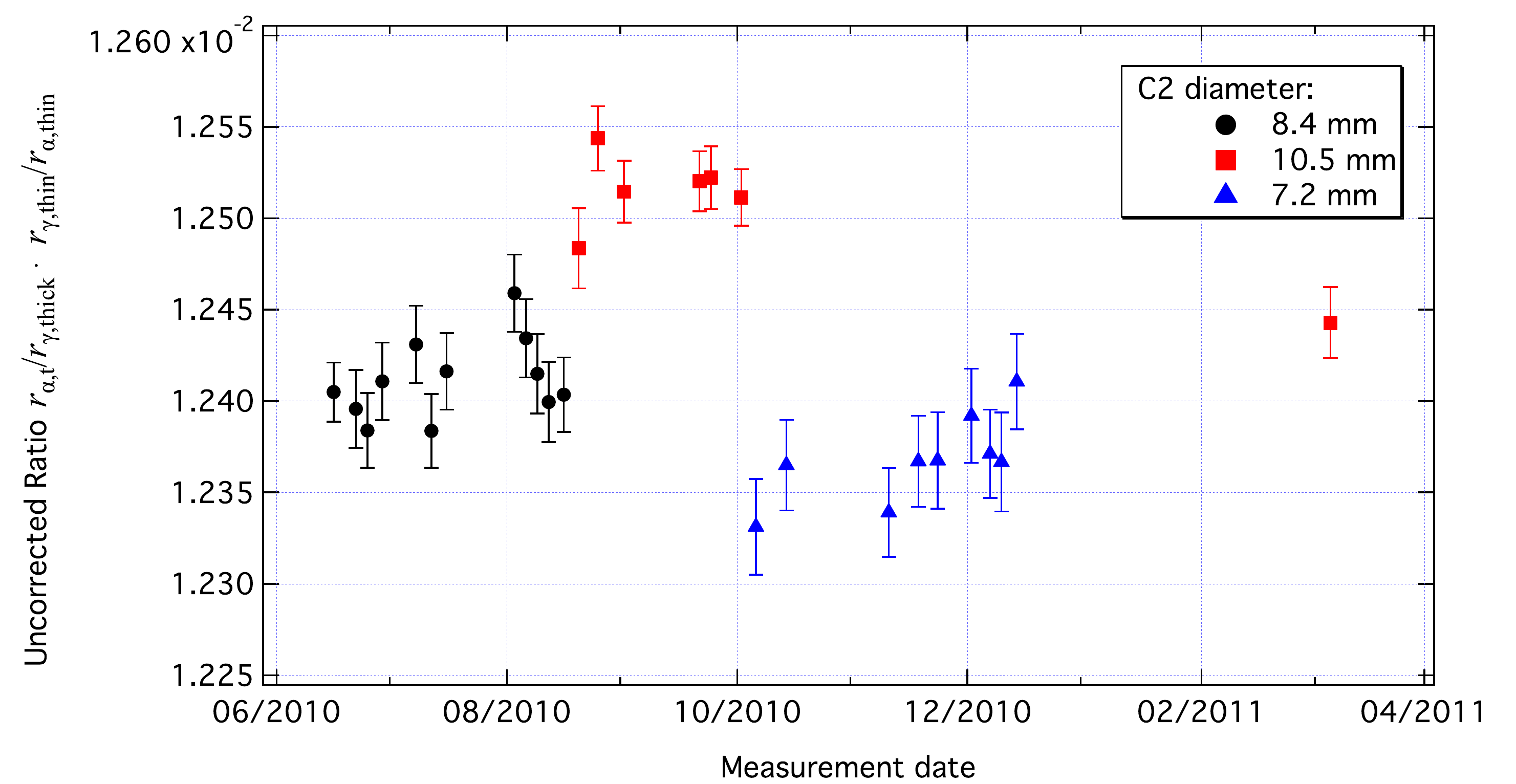}
	\caption{\label{fig:GFMAGData}Plot of the results of the analysis  for the three sizes of the C2 aperture. The error bars represent statistical uncertainty only.} 
\end{figure}

The statistical accumulation was performed at three beam sizes in order to investigate systematic effects related to beam size and total neutron rate. The beam size was varied by changing the diameter of the C2 aperture. Twelve calibrations were performed with C$2 = 8.38$\,mm, six with C$2 = 10.5$\,mm, and nine with C$2 = 7.2$\,mm. Figure~\ref{fig:GFMAGData} shows the data for the three sizes of the C2 aperture. To calculate the idealized $\epsilon_{0}(0,0)$ from the observed ratios requires careful determination of all systematic effects, as discussed in Section~\ref{sec:syst}. 

\section{Measurement of the neutron wavelength}
\label{sec:lambdasys}

\begin{figure}
  \centering
  \includegraphics[width=1\textwidth]{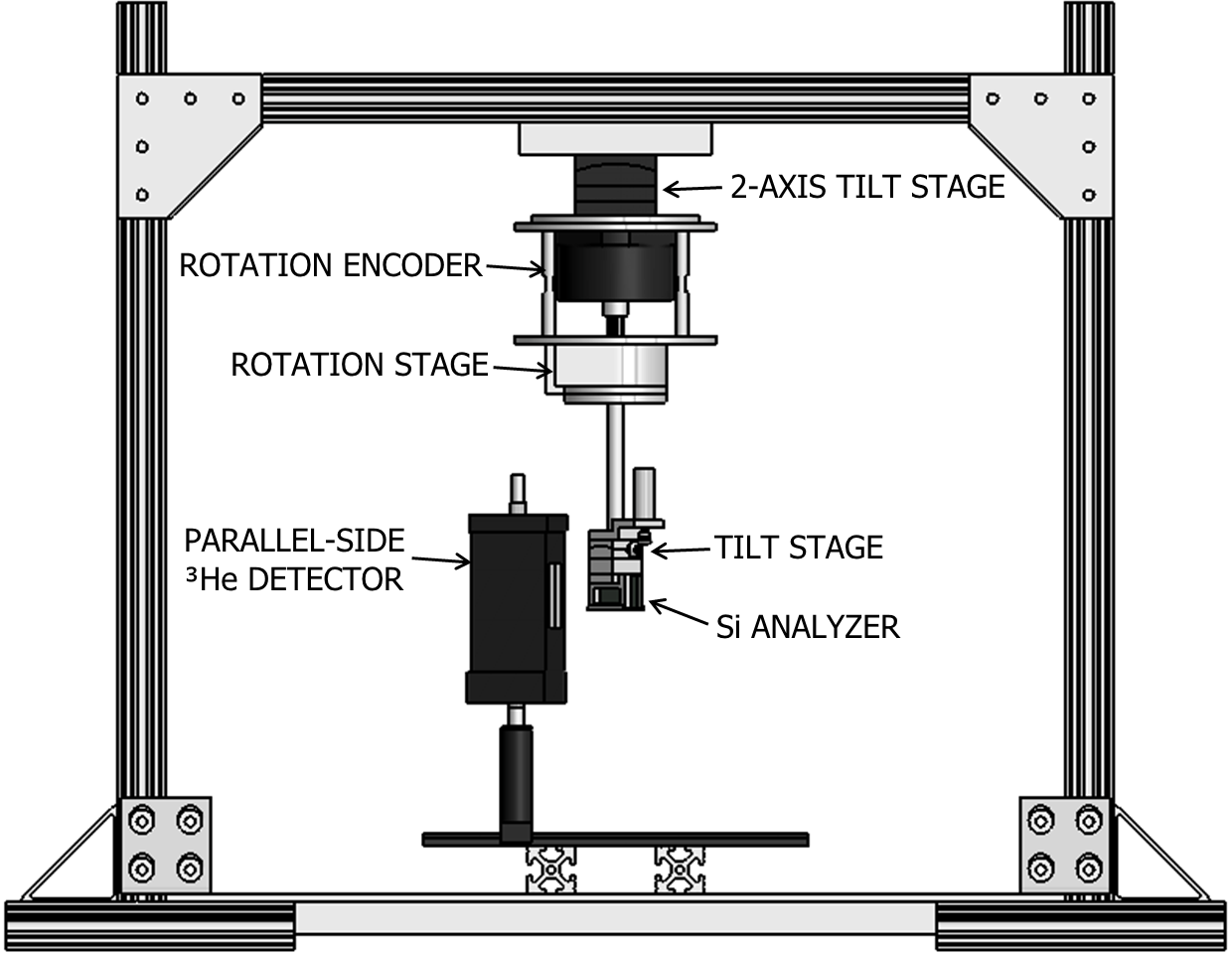}
  \caption{An illustration of the wavelength measuring apparatus installed on NG-6m.}
  \label{fig:NewLambdaSetup}
\end{figure}

An essential part of determining the efficiency of the neutron monitor is knowing the average wavelength (i.e., energy) of the neutron beam (see equation~\ref{eqn:epsiloneq}). The wavelength measurement and the efficiency measurements of the neutron monitor must be carried out on the same beam, and no changes to the wavelength can be permitted among the measurements.  The apparatus to measure the wavelength is illustrated in figure \ref{fig:NewLambdaSetup}; it was designed so that it could be inserted into the beamline without perturbing any of the critical beam-defining components.  The apparatus consisted of a manual two-axis tilt stage to adjust the crystal rotation axis direction, an encoder-rotation stage pair driven by a stepper motor for crystal rotation, and a one-axis tilt stage driven by a microstepper motor for tilting the crystal lattice planes.  The crystal positioning device was supported by a rigid frame.

To produce a monochromatic beam, polychromatic neutrons from the main beam are incident on a pyrolytic graphite crystal, and the direction and energy width of the reflected beam are determined by the lattice spacing ($d$) and orientation ($\theta$) of the crystal planes with respect to the incident beam.  For neutrons of wavelength $\lambda$, the Bragg condition is given by

\begin{equation}
\label{eqn:BraggsLaw}
n\lambda = 2d\sin\theta.
\end{equation}

\noindent Neutrons that satisfy the Bragg condition are reflected $2\theta$ from the main beam (roughly $90^{\circ}$).  The Bragg condition is met for approximately 0.5\,nm neutrons ($n = 1$) and, consequently, 0.25\,nm ($n = 2$) and higher order reflections.  Higher order components are strongly suppressed by a polycrystalline beryllium filter cooled to 77\,K that was placed in the beam.

\begin{figure}
  \centering
  \includegraphics[width=\textwidth]{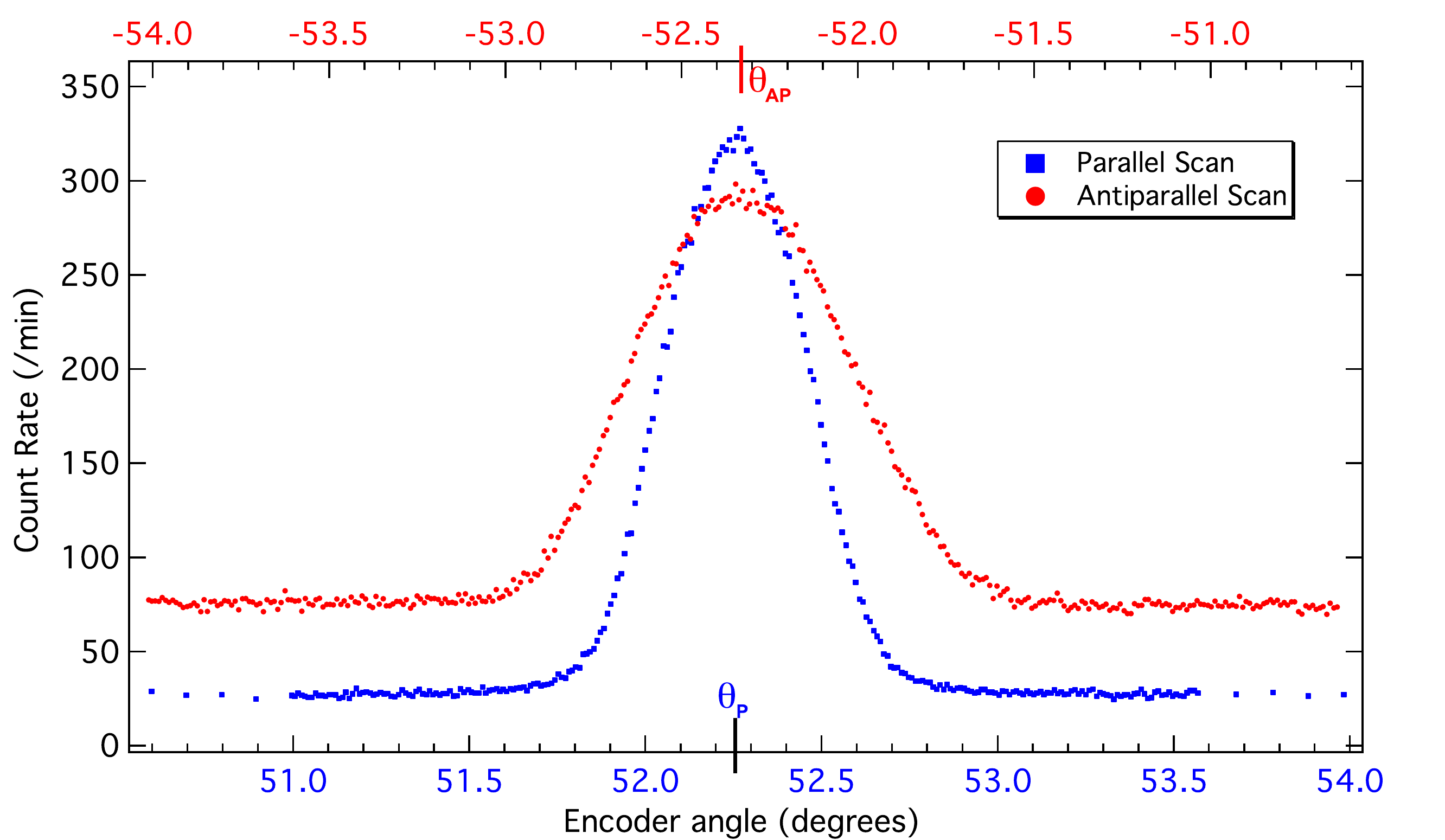}
  \caption{A typical rocking curve pair from the apparatus to measure the wavelength.}
  \label{fig:TypicalScan}
\end{figure}

\begin{figure}
  \centering
  \includegraphics[width=\textwidth]{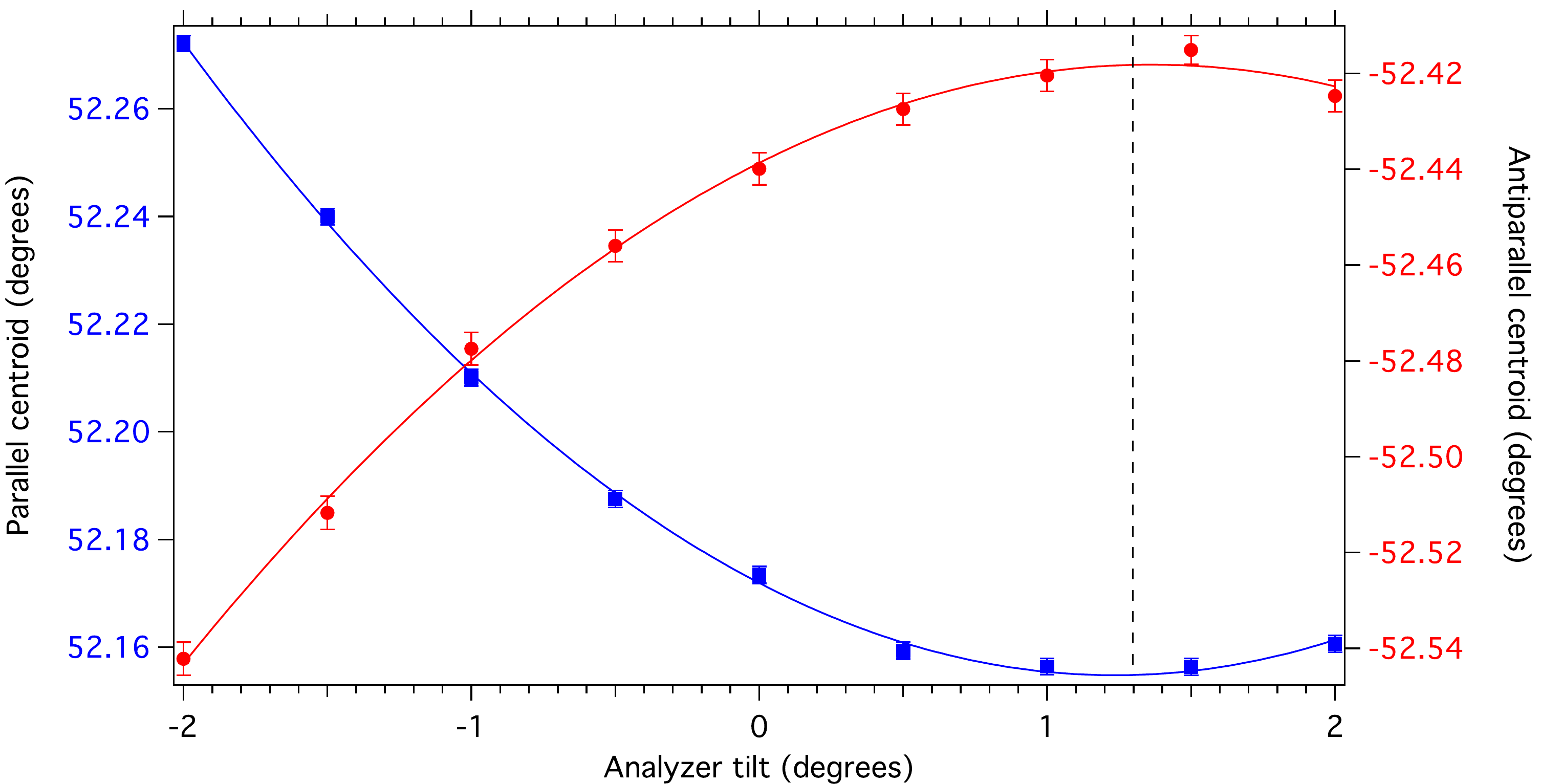}
  \caption{A tilt curve pair from the wavelength measuring apparatus. The solid lines are weighted fits to a parabola. The dashed line represents the tilt that best minimizes the parallel centroid and maximizes the antiparallel centroid. The error bars represent statistical uncertainty of the fit of the centroids.}
  \label{fig:TypicalTiltScan}
\end{figure}

The same principle used to extract the monochromatic beam was also used to measure its wavelength.  The neutron wavelength was measured by diffraction from the (111) planes of a silicon crystal analyzer in Laue geometry.  The reflected neutron intensity was measured in two ${}^{3}$He detectors positioned at the approximate angular position of the parallel ($\theta_{\rm{P}}$) and antiparallel ($\theta_{\rm{AP}}$) Bragg reflections.  The relative rotation angle of the analyzer crystal was measured by an encoder coupled to the crystal rotation shaft.  The centroids of the resulting plots of reflected intensity versus angular position, referred to as  ``rocking curves,''  determine $\theta_{\rm{P}}$ and $\theta_{\rm{AP}}$, as seen in figure \ref{fig:TypicalScan}.  In practice, the analyzer crystal planes were tilted an angle $\phi$ from normal to the beam, presenting a wider lattice spacing.  A tilting stage mounted to the crystal housing is used to deliberately tilt the crystal by an angle $\phi$ and rocking curves were performed to determine $\theta_{\rm{P}}$ and $\theta_{\rm{AP}}$ as a function of $\phi$.  The centroids of each tilt curve were fit to parabolas, as shown in figure \ref{fig:TypicalTiltScan}, and the minimum(maximum) of each parabola determines the true $\theta_{\rm{P}}$($\theta_{\rm{AP}}$), and thus \

\begin{equation}
\theta_{\rm{Bragg}} = \frac{\theta_{\rm{P}}-\theta_{\rm{AP}}}{2}.
\end{equation}

\begin{figure}[htbp]
  \centering
  \includegraphics[width=\textwidth]{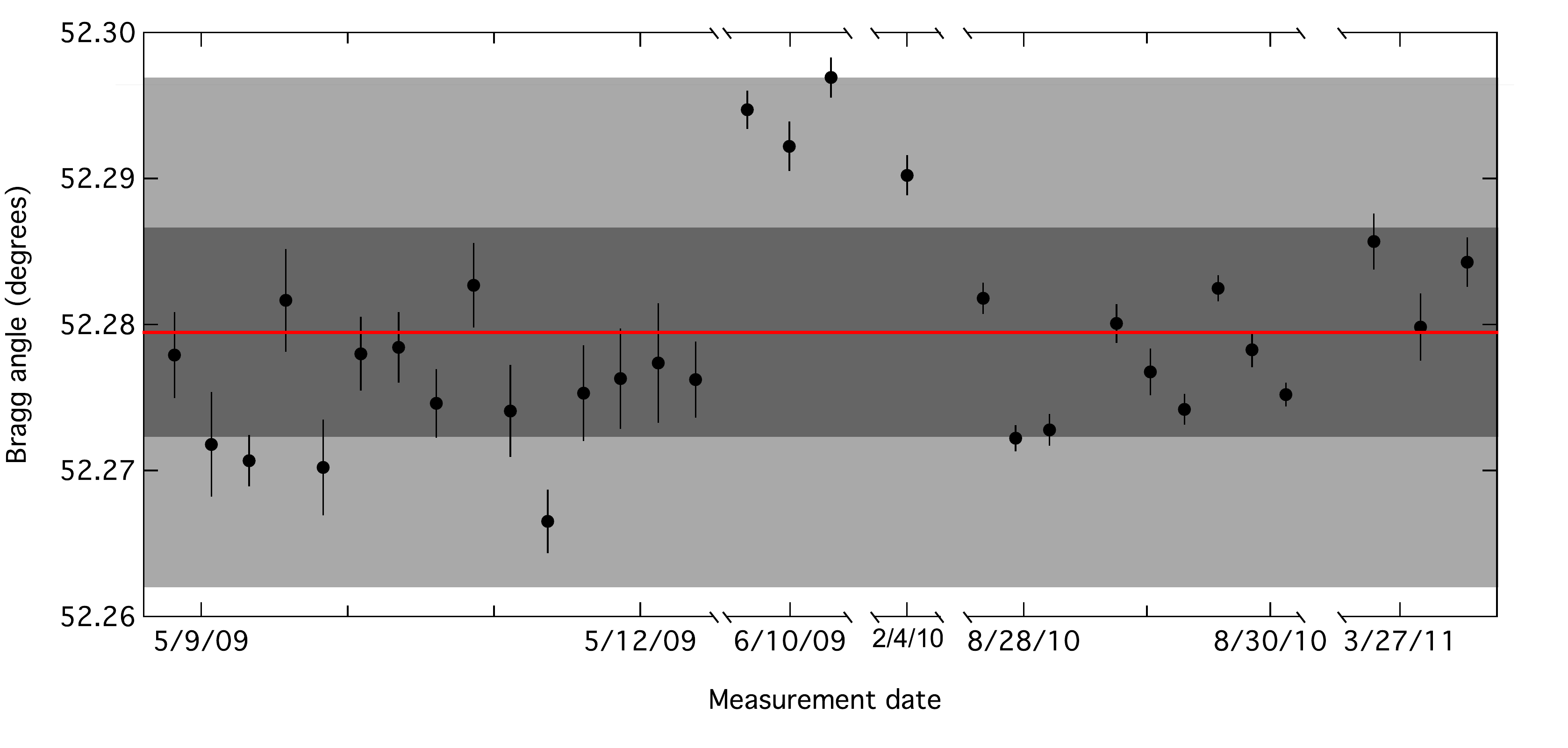}
  \caption{Plot of the measured $\theta_{\rm{Bragg}}$ angle.  The centroid of the data (solid red line) is determined from a weighted constant fit ($\chi^{2}/\nu = 23.2$), and the dark shaded band corresponds to the $\pm 1$-$\sigma$ uncertainty of that fit. The uncertainty used for the wavelength determination was conservatively chosen to include the entire data set and is denoted by the light shaded band.}
  \label{fig:newlambdadata}
\end{figure}

Figure \ref{fig:newlambdadata} gives a summary of the $\theta_{\rm{Bragg}}$ measurements.  The uncertainty in each point is statistical and comes from the determination of the centroids of the rocking curves. The May 2009 data was taken with the ${}^{3}$He detectors close to the analyzer crystal, increasing the scattered neutron background and thus the statistical uncertainty.  The origin of the excursion of $\theta_{\rm{Bragg}}$ in the June 2009 and February 2010 data is not known. As we found no a priori reason to exclude the data,  a weighted fit to the data was used to find the average $\theta_{\rm{Bragg}}$, and a conservative estimate of the uncertainty was made by using the maximum spread in the data.  This yields a value of $\theta_{\rm{Bragg}} = 52.279^{\circ} \pm 0.018^{\circ}$.  The silicon lattice spacing $a = (0.5431020504 \pm 0.000000087)$\,nm~\cite{Mohr06} is known to a relative precision of ${1.6}\times{10}^{-8}$, and thus the (111) spacing is given by

\begin{equation}
d_{111} = {{a}\over{\sqrt{(1^{2} + 1^{2} + {1}^2)}}} = (0.3135601150 \pm 0.0000000051)\, \rm{nm}.
\end{equation}

\noindent Substituting this value into equation \ref{eqn:BraggsLaw} yields $\lambda_{\rm{mono}} = (0.49605 \pm 0.00012)$ nm.

Because of the importance of the wavelength value, a separate measurement was performed to determine the contamination of the monochromatic beam by neutrons with wavelengths of $\lambda/2$.  Such neutrons will be detected in the flux monitor with half the probability of $\lambda$ neutrons but with equal probability by the Alpha-Gamma device in its totally-absorbing thick target mode.

The beryllium filter was removed and the angular position and intensity of the unfiltered $\lambda/2$ component of the beam were measured. Rocking curves were obtained in the  same manner as for the filtered beam, yielding a Bragg angle of $23.28^{\circ} \pm 0.02^{\circ}$ and a corresponding wavelength of $(0.2479 \pm 0.0002)$ nm. The unfiltered $\lambda/2$ rocking curves were each fit to a gaussian to establish the peak positions, widths, and heights.  Using the unfiltered $\lambda/2$ rocking curve fit values for width and position and keeping the ratio of parallel to antiparallel peak heights fixed, a fit is performed on the filtered rocking curves to determine the amplitude.  With the amplitude determined, we calculate the ratio of the area under the filtered $\lambda/2$ parallel rocking curve to the filtered $\lambda$ parallel rocking curve.  Simulation with McStas~\cite{McStas2014} has shown that, for our beam geometry, the ratio of the area under the $\lambda/2$ and $\lambda$ parallel rocking curves is equal to the ratio of the beam intensity for each component.  We find the ratio of $\lambda/2$ to $\lambda$ to be $(0.00065 \pm 0.00020)$, leading to a correction to the determined flux monitor efficiency of $(1.00032 \pm 0.00010)$




Other sources of systematic errors in the wavelength determination were considered and quantified. They include spatial variation of the reflectivity of the pyrolytic graphite used to produce the monochromatic beam, spatial variation of the mosaic orientation in the pyrolytic graphite, and misalignment of the silicon analyzer crystal (tilt error).  All of these effects were modeled in a Monte Carlo simulation that transported neutrons from the neutron guide tube all the way to the detector.  For expected sizes of these imperfections, the systematic uncertainties in the wavelength determination were negligible; the results of these investigations are detailed in Ref.~\cite{Coakley2003}.

\section{Determination of systematic effects and corrections}
\label{sec:syst}

Systematic effects must be accurately measured for each step of the calibration experiment.  There were six unique running configurations in the experiment, characterized by two gamma detectors (sensitive to different gamma scattering effects) and three beam sizes (sensitive to rate-dependent effects and solid angle).  In addition to configuration-dependent systematics, there were systematic effects common to each configuration.  In this section, each systematic effect is identified and expressed as a correction to the measured monitor efficiency $\epsilon$.

\subsection{Absolute alpha particle counting}
\label{sec:alphasys}

Ultimately, the determination of the neutron flux is premised upon the accurate determination of the absolute alpha-activity of the plutonium source.  The source activity was measured in a low-solid angle counting stack and includes several systematic corrections.  One must accurately determine the dead time of the counting system, the effect of alphas scattering off the source substrate, the solid angle of the counting stack, the perturbation to the solid angle for the extended source spot, and the tunneling of alphas through the aperture edge. The systematic effects are discussed here.

The approximate dead time is given by the full-width at half-maximum of the average signal amplifier output (${1.7}\,\mu$s) plus the time the SCA spends issuing a TTL pulse (0.5\,$\mu$s). The dead time was measured by a two source method~\cite{Yue2011,Moon1937,Janssen1994}, in which two measured rates were compared to the ratio of solid angles in the two setups.  For this measurement, a 1.5\,MBq ${}^{240}$Pu source was used at two different stack heights. Because of the high disintegration rate and the possibility of sputtering, a 30\,$\mu$g polyimide film was used to prevent contamination of the threaded spacers and detector surface.  The signal attenuation by the film was small~\cite{Gilliam08} but irrelevant because the two source method only depends on the observed count rates.  The dead time was determined to be $(2.205 \pm 0.050)$\,$\mu$s, which agrees very well with the pulse-width approximation.

Accurate determination of the stack solid angle was performed by coordinate measuring microscope and contacting metrology.  The diameter of a nickel-coated dimensionally-stable copper aperture was measured by a coordinate measuring machine. The result was $D_{\rm{ap}} = (25.765192 \pm 0.000240)$\,mm.  The source spot diameter was measured by a microscope giving a result of $D_{\rm{source}} = (2.992 \pm 0.008)$\,mm.  The source to aperture distance was also measured by the microscope yielding a result of $z_{\rm{ap}} = (87.4226 \pm 0.0015)$\,mm.

The emission of alpha particles is uniform over 4$\pi$, so backscattering from the Si backing is possible though infrequent.  The energy of these particles ranges from zero to the peak alpha energy.  Examination of the energy spectrum allows for the windows to be set in the flat region with few events from the noise tail and no forward-emitted (non-tunneled) alphas, leaving only backscattered alphas and tunneled alphas (see Figure~\ref{fig:Pualphaspec}).  By taking the difference between the two SCA counters, one can assess the number of backscattered alphas per channel and extrapolate the number of backscattered alphas in the peak.  Removal of backscattered alphas is a -0.04\,\% correction to the observed alpha count rate.

The flat region of the spectrum is a combination of Rutherford backscattered alphas and alphas that have tunneled through the aperture edge. The energy of a tunneling alpha-particle is a function of the aperture material and the distance the alpha has traversed in the aperture edge.  In the flat region where the energies are typically between 1\,MeV and 2\,MeV, this corresponds to a solid angle of $6\times10^{-5}$ for tunneled alphas.  Multiplying this solid angle by the source emission rate and comparing with observed rates in this energy range, we calculate that 63\,\% of the alphas are tunneled alphas.  These counts are attributed to the detected total, whereas the Rutherford scattered alphas, or remaining counts, must be removed.  By including the tunneled alphas, the effective aperture diameter is increased to $(25.76694 \pm 0.00058)$\,mm.

Taking into account the dead time, Rutherford backscattered alphas, and tunneled alphas, we find the observed alpha rate is $(125.740 \pm 0.023)$ s${}^{-1}$.  From the measured dimensions, the effective solid angle of the stack for these alphas is $\Omega_{\rm{stack}} = 0.0053415 \pm 0.0000004$, thus yielding an absolute activity for the source $R_{\rm{Pu}} = (23540.4 \pm 4.6)$\,s${}^{-1}$.

As discussed, the calibrated source is used in the Alpha-Gamma device as the first step in determining $\Omega_{\rm{AG}}$ for each beam size.  By dividing the measured alpha rate by the known absolute source activity, one determines the effective detection solid angle for the alpha detector.  This solid angle is perturbed by the source spot size, tunneling through the aperture edges, and Rutherford backscattering. The alpha detector aperture is measured on the coordinate measuring microscope, resulting in $D_{\rm{AG}} = (27.598 \pm 0.006)$\,mm.  The aperture is made of aluminum, and the effective diameter for Pu alphas is $(27.603 \pm 0.006)$\,mm.  The observed rate of alphas is $r_{\rm{Pu,AG}} = (168.420 \pm 0.017)$\,s${}^{-1}$.  From the observed alpha rate, the known absolute activity, and the known aperture properties, we calculate the source to aperture distance to be $z_{\rm{AG}} = (80.696 \pm 0.020)$\,mm.  This allows us to calculate $\Omega_{\rm{AG}}(x,y)$ for any point on the deposit surface for alphas of any energy.

\subsection{Thin \Bt\ target corrections}
\label{sec:thinsyst}

The 1.4\,MeV to 1.8\,MeV alpha particles emitted from neutron capture on \Bt\ are detected in the Alpha-Gamma alpha detector to determine the neutron absorption rate in the thin \Bt\ target.  The neutron rate $r_{\rm{n,thin}}$ is given approximately by equation \ref{eqn:rnthin} and, more accurately, by taking into account the proper average detection solid angle $\bar{\Omega}_{\rm{AG}}$

\begin{equation}
r_{\rm{n,\rm{thin}}} = \frac{r_{\alpha,\rm{thin}}}{\bar{\Omega}_{\rm{AG}}}.
\end{equation}

\begin{figure}
	\centering
\includegraphics[width=\textwidth]{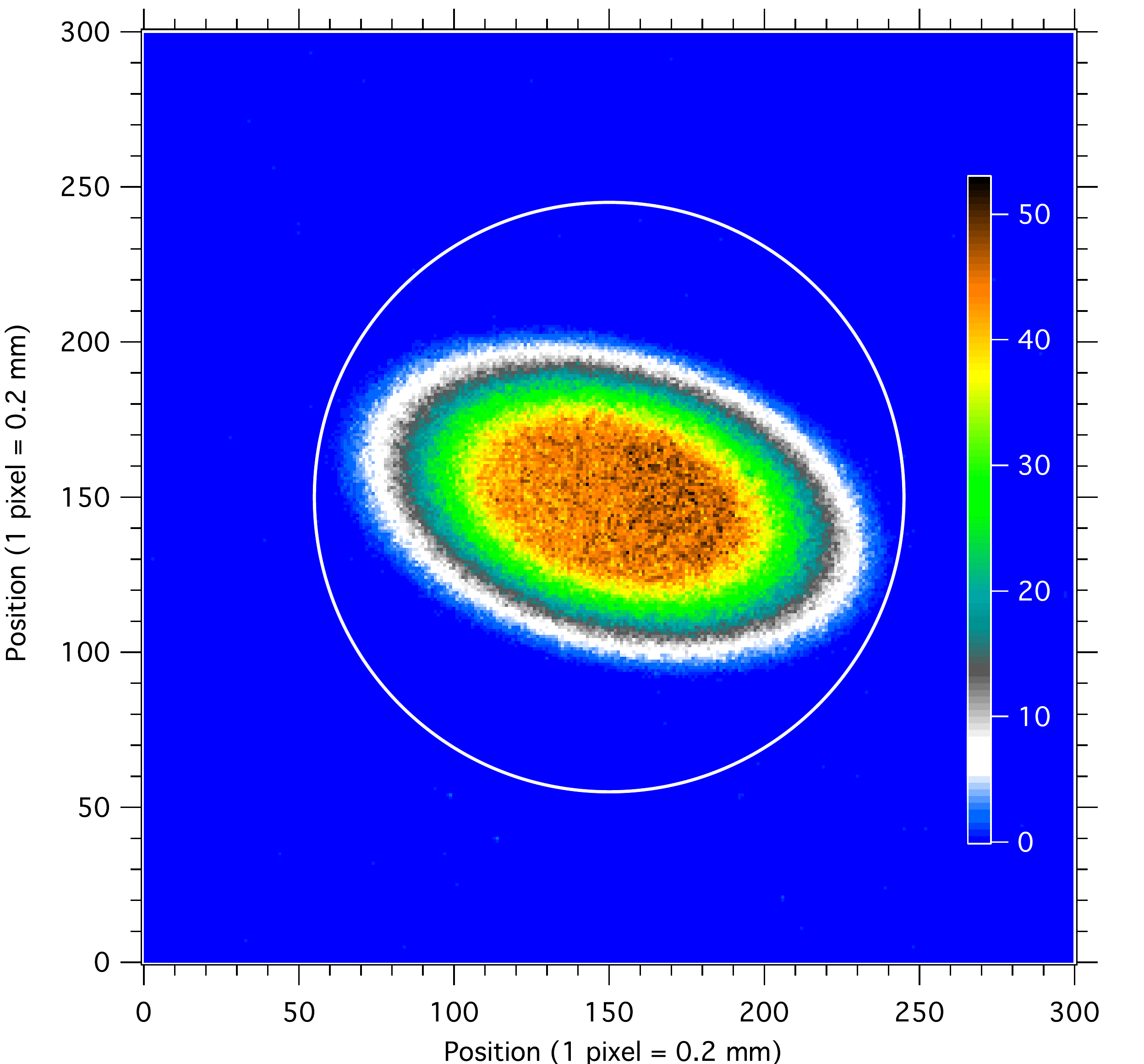}
	\caption{Beam images were acquired at the Alpha-Gamma deposit location with C1 = 15\,mm and C2 = 7.2, 8.38, and 10.5\,mm. The image for the largest collimation of C2 = 10.5\,mm is shown here; the color scale represents linear intensity, and the image orientation is arbitrary. The white circle corresponds to the edge of the active area of the thin \Bt\ deposit.}
  \label{fig:AGBeamImages}
\end{figure}

The average solid angle can be calculated from $\Omega_{\rm{AG}}(x,y)$, the density profile of the deposit $\rho(x,y)$ and the intensity profile of the beam striking the target $I(x,y)$

\begin{equation}
\bar{\Omega}_{\rm{AG}} = \frac{1}{C}\int\int{\Omega_{\rm{AG}}(x,y)\rho(x,y)I(x,y)dxdy},
\label{eqn:OmegaCalc}
\end{equation}

\noindent where $C = \int\int{\rho(x,y)I(x,y)}dxdy$.  The density profile of the deposit is known from the characterization described previously (equation \ref{eqn:rhor}).  The intensity profile of the beam is measured by replacing the thin \Bt\ target with a dysprosium deposit.  Approximately 20\,\% of natural Dy is $^{164}$Dy, which has a large thermal neutron cross section.  The resultant isotope (${}^{165}$Dy) beta decays with a 2.3 h half-life.  The irradiated deposit is exposed to a radiation-sensitive Fuji image plate, which is then read out in a plate reader.  The measured photo-stimulable luminescence is directly proportional to the intensity of the incident radiation over a wide dynamic range.  A 1\,mm diameter hole marks the center of the dysprosium deposit, providing the origin of the deposit coordinate system, and thus $I(x,y)$.  The uncertainty is determined by performing multiple beam images and using the average determined $\bar{\Omega}_{\rm{AG}}$.  We estimate the uncertainty to be the standard deviation of the measured solid angles divided by the square root of the number of images taken.  The image for the largest collimation is shown in figure \ref{fig:AGBeamImages}, and the results for each collimation are given in table \ref{tab:AGAOmega}. 

\begin{table}
\caption{\label{tab:AGAOmega} Solid angle parameters for the three beam sizes used in the measurement. The first column gives the diameter of the upstream/downstream apertures in units of mm; the second column is the number of Dy images that were acquired; the third column is the solid angle $\bar{\Omega}_{\rm{AG}}$ of the Alpha-Gamma alpha detector; and the last column is the resulting efficiency correction.}
\centering
\begin{tabular}{cccc}
\hline\hline\noalign{\smallskip}
Collimation & Number of images & $\bar{\Omega}_{\rm{AG}}$ & $\epsilon$ correction $\left(\bar{\Omega}_{\rm{AG}}/\Omega_{\rm{AG}}(0,0)\right)$\\
\noalign{\smallskip}\hline\noalign{\smallskip}
15/7 & 2 & 0.0070985 $\pm$ 0.0000003	& 0.99215 $\pm$ 0.00004	\\
15/8 & 3 & 0.0070810 $\pm$ 0.0000003	& 0.98970 $\pm$ 0.00005	\\
15/10 & 5 & 0.0070538 $\pm$ 0.0000004	& 0.98590 $\pm$ 0.00005	\\
\noalign{\smallskip}\hline\hline
\end{tabular}
\end{table}

The remaining systematic effects in the thin \Bt\ target are gamma-ray counting effects.  The silicon backing on the thin target causes gamma-ray production by neutron absorption and gamma-ray attenuation as the \Bt\ capture gammas travel to the top gamma detector.  Only approximately 1\,\% of neutrons that impinge on the thin target will be absorbed by \Bt.  A small fraction of the remaining neutrons will interact with the Si backing wafer and several capture gamma-rays are produced.  These lines are of higher energy ($\geq 1.5$\,MeV) than the 478\,keV boron capture gamma but can Compton scatter in the germanium crystal and incompletely deposit their energy.  This background is not removed by measuring the thin target gamma background with the upstream ${}^{6}$Li flag blocking the beam.  Instead, the Si gamma background in the 478\,keV signal region is determined by long runs with a Si blank target instead of the usual thin target.  This background is a function of incident neutron flux, so the measured gamma rate must be divided by neutron flux.  Because a silicon blank is used as the Alpha-Gamma target, the only choice for neutron flux assessment is the neutron flux monitor.  Thus, the relevant experimental quantity is $\gamma/\rm{FM}$ which depends on C2 and gamma detector efficiency.  A number of measurements of the silicon gamma-ray background were performed over the course of the calibration data.  The correction for each set of data is shown in table \ref{tab:ThinTargetSiGammas}.

\begin{table}
\caption{\label{tab:ThinTargetSiGammas}Silicon $\gamma$ as a fraction of the total measured 478\,keV signal for each running configuration. ``Detector-collimation" refers to the top (T) or bottom (B) Ge detector and the diameter of the collimation in millimeters.}
\centering
\begin{tabular}{ccc}
\hline\hline\noalign{\smallskip}
Detector-collimation & Si $\gamma$ fraction ($ \times{10}^{-2}$) & $\epsilon$ correction \\
\noalign{\smallskip}\hline\noalign{\smallskip}
T-7 & 1.1715 $\pm$ 0.0009	& 0.988284 $\pm$ 0.000009	\\
T-8 & 1.1995 $\pm$ 0.0006	& 0.988005 $\pm$ 0.000006	\\
T-10 & 1.2044 $\pm$ 0.0005	& 0.987956 $\pm$ 0.000005	\\
B-7 & 1.3235 $\pm$ 0.0018	& 0.986765 $\pm$ 0.000018	\\
B-8 & 1.2145 $\pm$ 0.0007	& 0.987851 $\pm$ 0.000007	\\
B-10 & 1.2450 $\pm$ 0.0007	& 0.987550 $\pm$ 0.000007	\\
\noalign{\smallskip}\hline\hline
\end{tabular}
\end{table}

\begin{figure}[htbp]
  \centering
  \includegraphics[width=\textwidth]{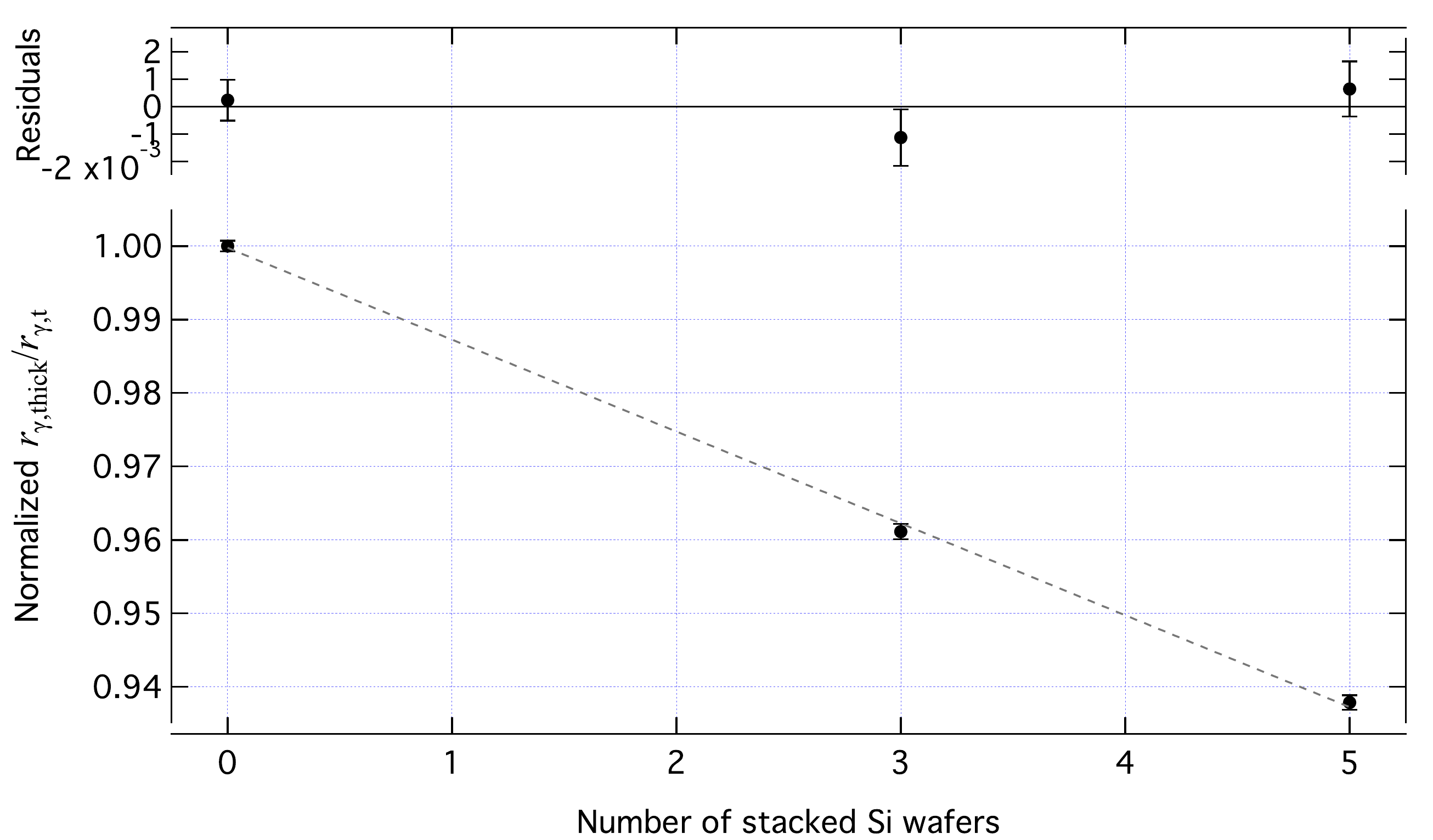}
  \caption{A plot of the ratio $\frac{r_{\gamma,\rm{thick}}}{r_{\alpha,t}}$ versus number of silicon wafers stacked behind the \Bt\ target.  The dashed line is a linear fit to the data; the residuals are shown in the upper plot. The error bars represent statistical uncertainty only.}
  \label{fig:SiGammaScatter}
\end{figure}

In addition to neutron interaction in the Si substrate, gamma interactions also occur.  \Bt\ capture gamma rays originate from the front face of the target and must travel through the 0.4\,mm silicon backing to reach the top gamma detector.  A simple calculation using XCOM cross sections~\cite{XCOM} and a beam simulation to determine average gamma path length in the material shows that approximately 1\,\% of the gamma rays scatter in the Si backing.  A measurement is needed to determine the corrections to sufficient precision.  In this measurement, the gamma rate in the top detector is measured with some number of Si backings behind the target deposit, and normalized to the beam rate ($\frac{r_{\gamma,\rm{thick}}}{r_{\alpha,t}}$). Measurements were performed with 0, 3, and 5 silicon deposits.  The slope is determined from a linear fit to the data, shown in figure \ref{fig:SiGammaScatter}, establishing the gamma attenuation per unit length. To correct for this effect, a multiplicative correction factor of $1.01298 \pm 0.00025$ is applied to the top detector data.

\subsection{Thick ${}^{10}$B${}_{4}$C target corrections}
\label{sec:thicksys}

A thin (0.32\,mm), self-supporting target of highly enriched boron carbide (${}^{10}$B${}_{4}$C, 98\,\% enrichment) is sufficiently thick to stop a beam of cold neutrons to better than 0.9999 absorption.  For the calibration to be accurate, it is necessary to determine scattering and reaction channels that do not result in the absorption of neutrons by \Bt.  Gamma losses not common to both the thin and thick targets must also be corrected appropriately.

\begin{figure}[htbp]
  \centering
  \includegraphics[width=\textwidth]{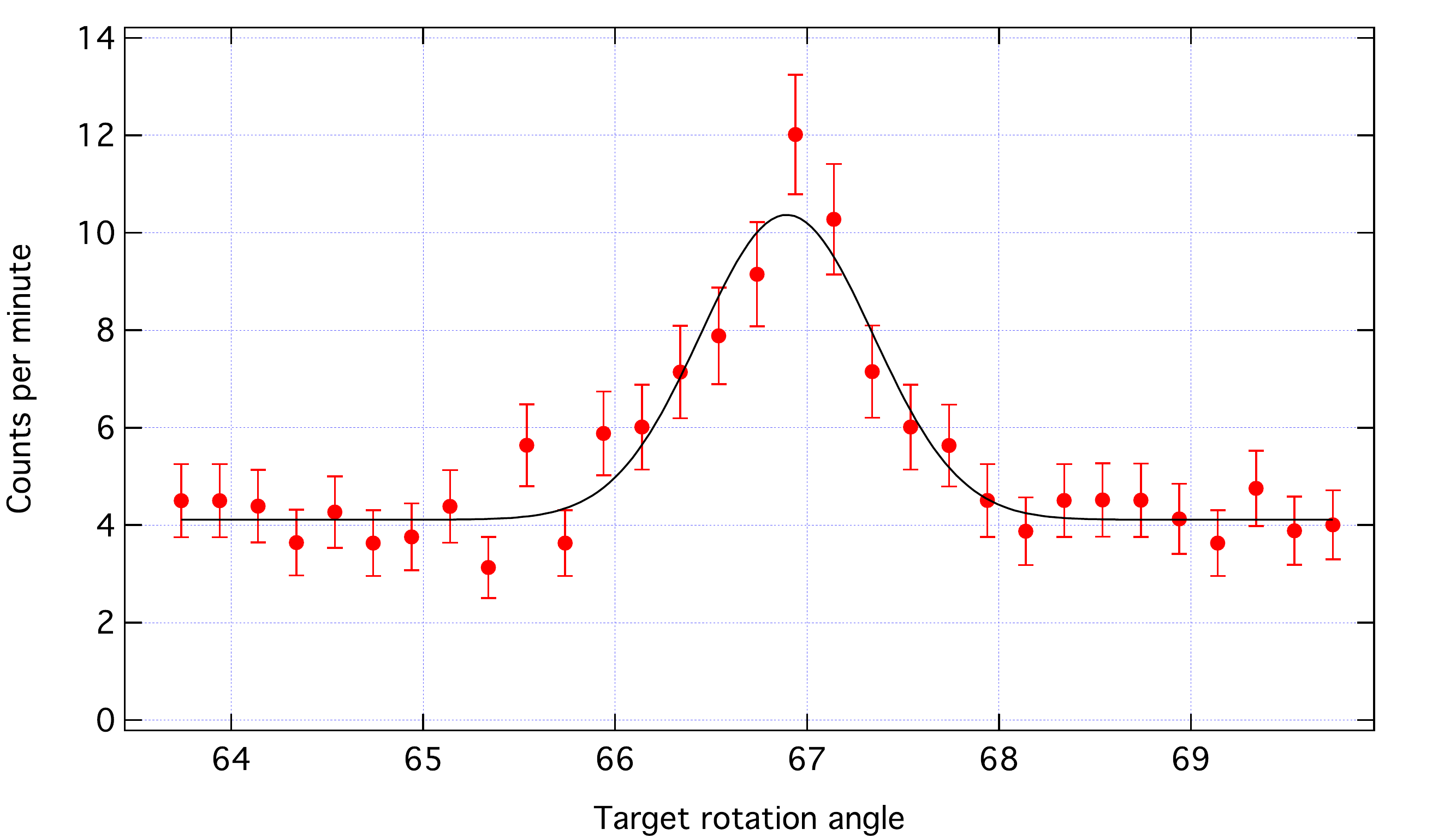}
  \caption{Plot of the measurement of the (101) reflection angle in the boron carbide target.The error bars represent statistical uncertainty only. The solid line is a fit to a Gaussian and flat background. The error bars represent statistical uncertainty only.}
  \label{fig:SPINSdata}
\end{figure}

Neutron scattering from the thick target can take place in three ways: coherent scattering from crystalline regions, scattering from surface features, and incoherent scattering.  Boron carbide is a ceramic and is likely polycrystalline, so Bragg scattering from the material is possible though the number of neutrons lost to the effect is very small.  Coherent scattering was assessed by powder diffraction techniques.  Scattering from a thicker thick target was measured on the SPINS apparatus at the NCNR.  One Bragg peak consistent with the (101) reflection was measured (figure \ref{fig:SPINSdata}).  The scattered fraction into this peak is approximately $2\times{10}^{-7}$, making the effect completely negligible.

Two additional techniques were used to assess the backscattered fraction.  In one, the dysprosium disc was placed near the alpha detector in the Alpha-Gamma device and the 0.321\,mm ${}^{10}$B${}_{4}$C target was loaded into the deposit holder.  The beam was turned on for an hour, and Dy was exposed to the Fuji plate for 15 minutes. No counts were seen in the Dy deposit image after background subtraction.  A second method used two approximately ${100}$ $\mu$g/cm${}^{2}$ thin \Bt\ deposits, one placed in the neutron monitor and the other placed in the Alpha-Gamma device.  The neutron beam was turned on and the ratio of the alpha rates observed in the two detectors was recorded.  A second run was performed with the thick, 0.321\,mm ${}^{10}$B${}_{4}$C target placed directly behind the thin \Bt\ deposit in the Alpha-Gamma device.  If any neutrons backscatter from the thick target, it would enhance the observed alpha rate in the Alpha-Gamma device.  A fractional count rate change at the level of $(-4 \pm 6)\times{10}^{-4}$, consistent with zero, was observed.  

The incoherent scattering from the thick target is calculable.  As seen in table \ref{tab:BandCcrosssections}, absorption and scattering from ${}^{11}$B, ${}^{12}$C, and ${}^{13}$C are negligible.  Only the \Bt\ incoherent cross section of 3\,b is important for the calculation.  For ${0.5}$ nm neutrons, the absorption cross section is $(10580 \pm 25)$\,b.  A simulation is performed to determine the number of neutrons that scatter from the target face but do not ultimately absorb.  We find that 0.006\,\% of the neutrons incident on the thick target will scatter, leading to a multiplicative correction to the measured $\epsilon$ of $0.9999405 \pm 0.0000003$.

\begin{table}
\caption{\label{tab:BandCcrosssections}Incoherent scattering and absorption cross sections for isotopes of boron and carbon~\cite{Sears1992,NCNR2016}.}
\centering
\begin{tabular}{cccc}
\hline\hline\noalign{\smallskip}
Isotope& Relative concentration & $\sigma_{inc}$ (b) & $\sigma_{abs}$ (b) \\	
\noalign{\smallskip}\hline\noalign{\smallskip}
${}^{10}$B 	& 0.784 	&  3 		& 10580		\\
${}^{11}$B 	& 0.016 	& 0.21 	& 0.015		\\
${}^{12}$C 	& 0.198 	& 0 		& 0.00974		\\
${}^{13}$C 	& 0.0022 	& 0.034 	& 0.00378		\\
\noalign{\smallskip}\hline\hline
\end{tabular}
\end{table}

\begin{figure}[htbp]
  \centering
  \includegraphics[width=0.65\textwidth]{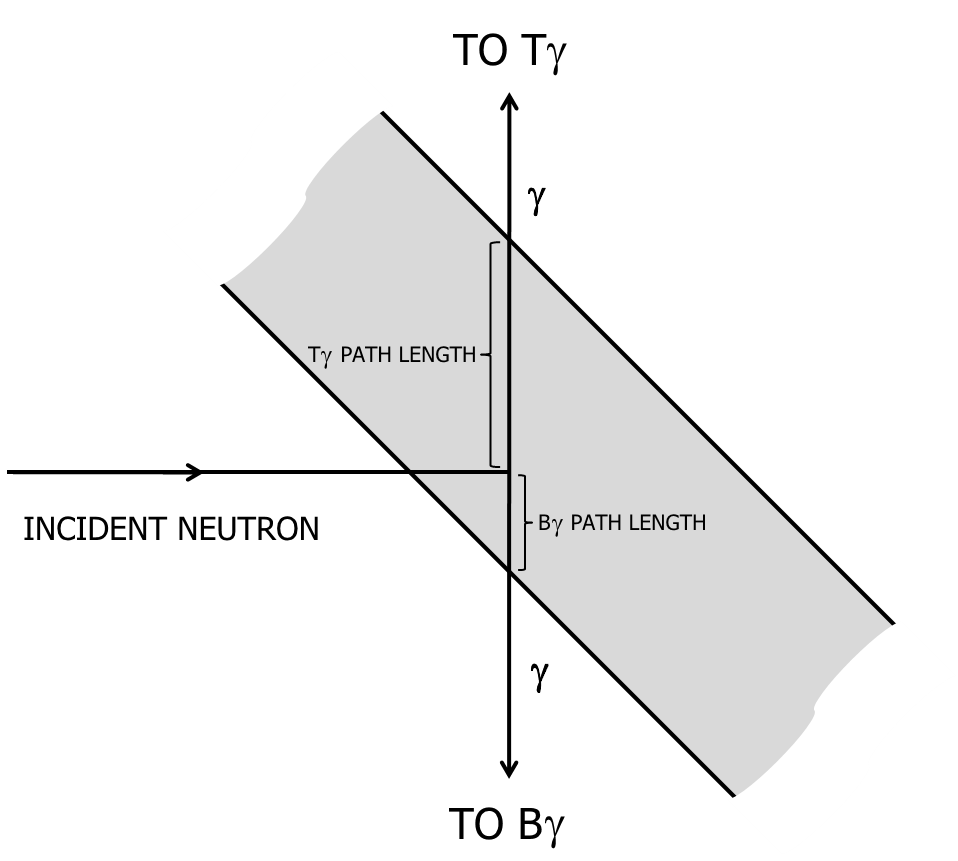}
  \caption{An illustration of the different path lengths through the thick target.  The travel distance through the target to the bottom detector is greatly exaggerated.}
  \label{fig:ThickPathLengthCartoon}
\end{figure}

Neutrons are absorbed at a variety of depths within the thick target (figure \ref{fig:ThickPathLengthCartoon}).  This leads to an average distance a capture gamma ray must travel in the material to reach one of the two gamma detectors.  A simulation of neutrons incident on the thick target takes into account the average extinction length and records the final position of each neutron.  The path length from the absorption sites to each of the detectors is recorded.  The average path length to the bottom detector is 0.036\,mm and the average path length to the top detector is 0.547\,mm.  The detection geometry cannot easily be changed to measure the attenuation to the bottom detector, but a measurement of the gamma attenuation to the top detector is straightforward.  

\begin{figure}[htbp]
  \centering
  \includegraphics[width=\textwidth]{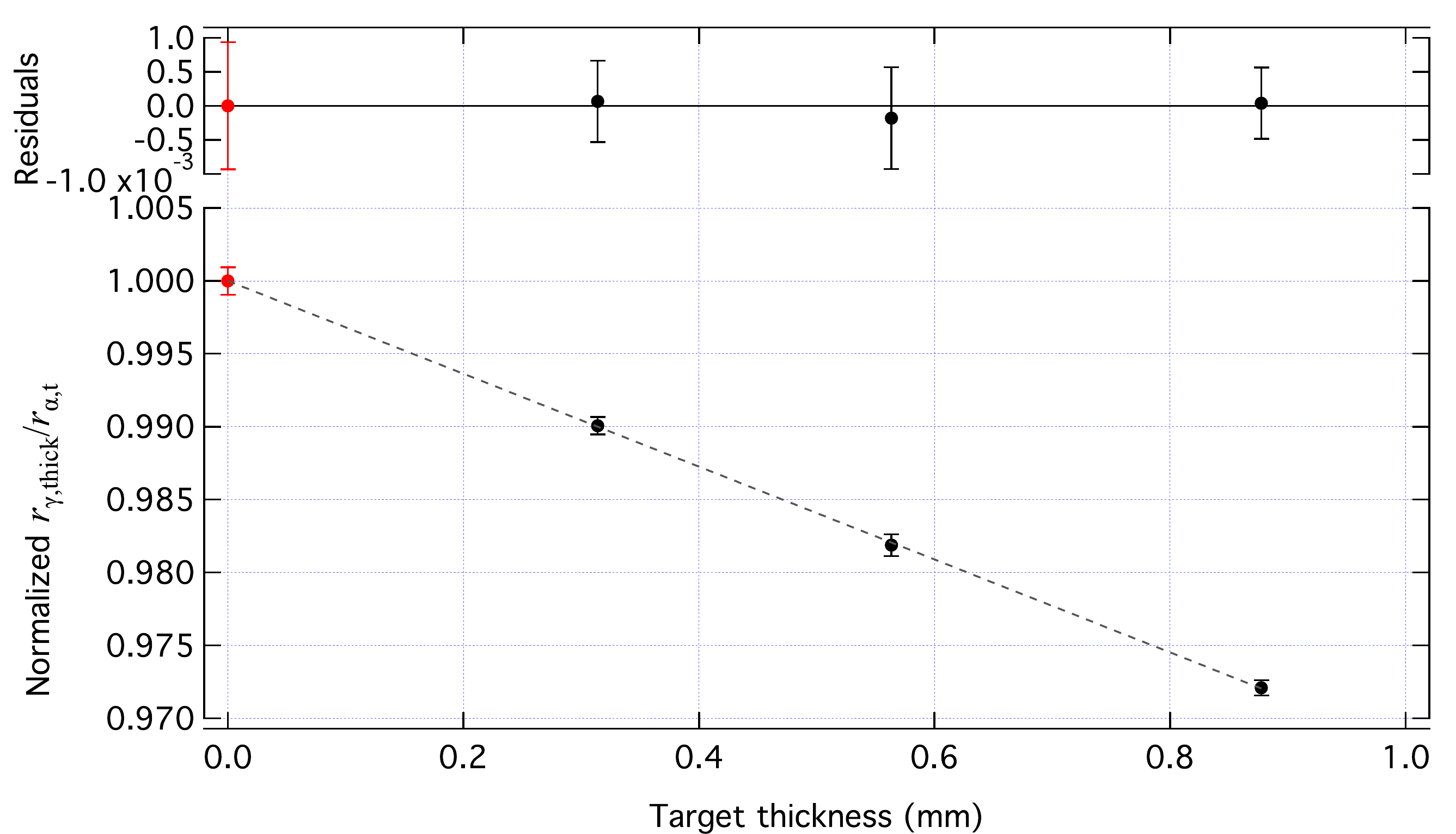}
  \caption{A plot of the ratio $\frac{r_{\gamma,\rm{thick}}}{r_{\alpha,t}}$ for three thicknesses of B${}_{4}$C: 0.321\,mm, 0.571\,mm, and 0.892\,mm. The black points are measured values, and the point for the largest thickness is the sum of the first two. The dashed line is a linear fit to those three data points, and the residuals are shown in the upper plot.  The red point is the result of the fit at 0 thickness, and the y-axis was normalized to make that point be 1. This was done to make the magnitude of the correction factor obvious. The error bars represent statistical uncertainty only.} 
  \label{fig:B4CGammaScatter}
\end{figure}

Two thick ${}^{10}$B${}_{4}$C targets were acquired for this experiment.  The second target is identical in composition but is 0.571 mm thick.  By measuring the beam-normalized gamma rate ($\frac{r_{\gamma,\rm{thick}}}{r_{\alpha,t}}$) in the top detector with the 0.321\,mm target, the 0.571\,mm target, and a stack of both targets, one can extract the gamma attenuation in ${}^{10}$B${}_{4}$C per unit length, as shown in figure \ref{fig:B4CGammaScatter}.  From the experimentally determined attenuation for the top detector and the simulated gamma distance in the material for the top and bottom detectors, the attenuation to the bottom detector is determined.  For our calibration data, the thinner thick target was used.  The results are given in table \ref{tab:B4CGammaAttenuation}.

\begin{table}
\caption{\label{tab:B4CGammaAttenuation}Lost 478\,keV $\gamma$ due to scattering in B${}_{4}$C target.}
\centering
\begin{tabular}{ccc}
\hline\hline\noalign{\smallskip}
Gamma Detector& \,\% scattered& $\epsilon^{\rm{AG}}_{0}$ correction\\
\noalign{\smallskip}\hline\noalign{\smallskip}
Top 		& 0.990 $\pm$ 0.048 	& 0.9901 $\pm$ 0.0005	\\
Bottom 	& 0.069 $\pm$ 0.003	& 0.99931 $\pm$ 0.00003	\\
\noalign{\smallskip}\hline\hline
\end{tabular}
\end{table}

The narrow collimation requirements of the experiment as well as the inherently low rate of the monochromatic beam kept signal rates at levels where dead time corrections are negligible for all signals but the thick target gamma signal.  The signal rate in the 478\,keV region is not particularly high (typically 300 s${}^{-1}$), but one must consider the entire detected gamma rate of approximately ${1000}$\,s${}^{-1}$, which is high enough to cause about ${1}$\,\% counting losses.  A pulser method is used to determine the detector dead time~\cite{Anders1969}.  A precision 25\,Hz fixed-interval pulser was input into the test input of both gamma detectors.  The pulser peak location was chosen to be in a region of very low background.  SCA windows were placed the same number of channels away from the edge of the pulser peak as they were around the boron photopeak.  This distance is important because events that just barely pile-up could merely shift inside the peak region and still be counted.  It is then a straightforward matter of taking the counts in the pulser peak region, subtracting the background, and comparing that to the known 25\,s${}^{-1}$ rate of the pulser.  The dead time is significant in all configurations: from 0.4\,\% for the 7.2\,mm top gamma detector data to 1.3\,\% for the 10.5\,mm bottom gamma detector data.  The rate is roughly ten times higher than that of the ${}^{239}$Pu source, and the amplifier dead time is about three times larger.  A summary of the results is found in table \ref{tab:ThickTargetDeadTimes}.

Neutrons absorbed via the \(^{10}\)B\(\left(\textrm{n},\gamma\right)^{11}\)B reaction emit gamma rays of much higher energy (approximately 4\,MeV to 10\,MeV) than the 478\,keV signal gamma. While these can Compton scatter just like the Si capture gammas, they are present in equal fraction during the thick and thin target data taking; therefore, Compton scattered radiative capture gammas which deposit energy within the 478\,keV window are naturally incorporated into the gamma detector calibration. What is not accounted for is the
neutron loss due to this channel. The correction is calculated in a straightforward manner using the measured cross section for the \(^{10}\)B\(\left(\textrm{n},\gamma\right)^{11}\)B
\cite{Bartholomew1957,Kok1986,Firestone2016}.  Taking a weighted mean of the cross section and comparing it to the \(^{10}\)B\(\left(\textrm{n},\gamma\right)^{7}\)Li cross section, it is seen that radiative capture absorbs $0.00010 \pm 0.000004$ of the neutrons. This results in a multiplicative correction to the measured flux monitor efficiency of \(0.99990 \pm 0.000004\).

\begin{table}
\caption{\label{tab:ThickTargetDeadTimes}Gamma detector dead time corrections for the thick target with each collimation scheme.}
\centering
\begin{tabular}{cc}
\hline\hline\noalign{\smallskip}
Detector - C2& $\epsilon^{\rm{AG}}_{0}$ correction\\
\noalign{\smallskip}\hline\noalign{\smallskip}
T-7 	& 0.99533 $\pm$ 0.00003	\\
T-8 	& 0.99339 $\pm$ 0.00002	\\
T-10	& 0.99006 $\pm$ 0.00004	\\
B-7	& 0.99441 $\pm$ 0.00010	\\
B-8	& 0.99201 $\pm$ 0.00007	\\
B-10	& 0.98800 $\pm$ 0.00011	\\
\noalign{\smallskip}\hline\hline
\end{tabular}
\end{table}

\subsection{Flux monitor corrections}
\label{sec:FMcorr}

\begin{figure}[htbp]
  \centering
  \includegraphics[width=\textwidth]{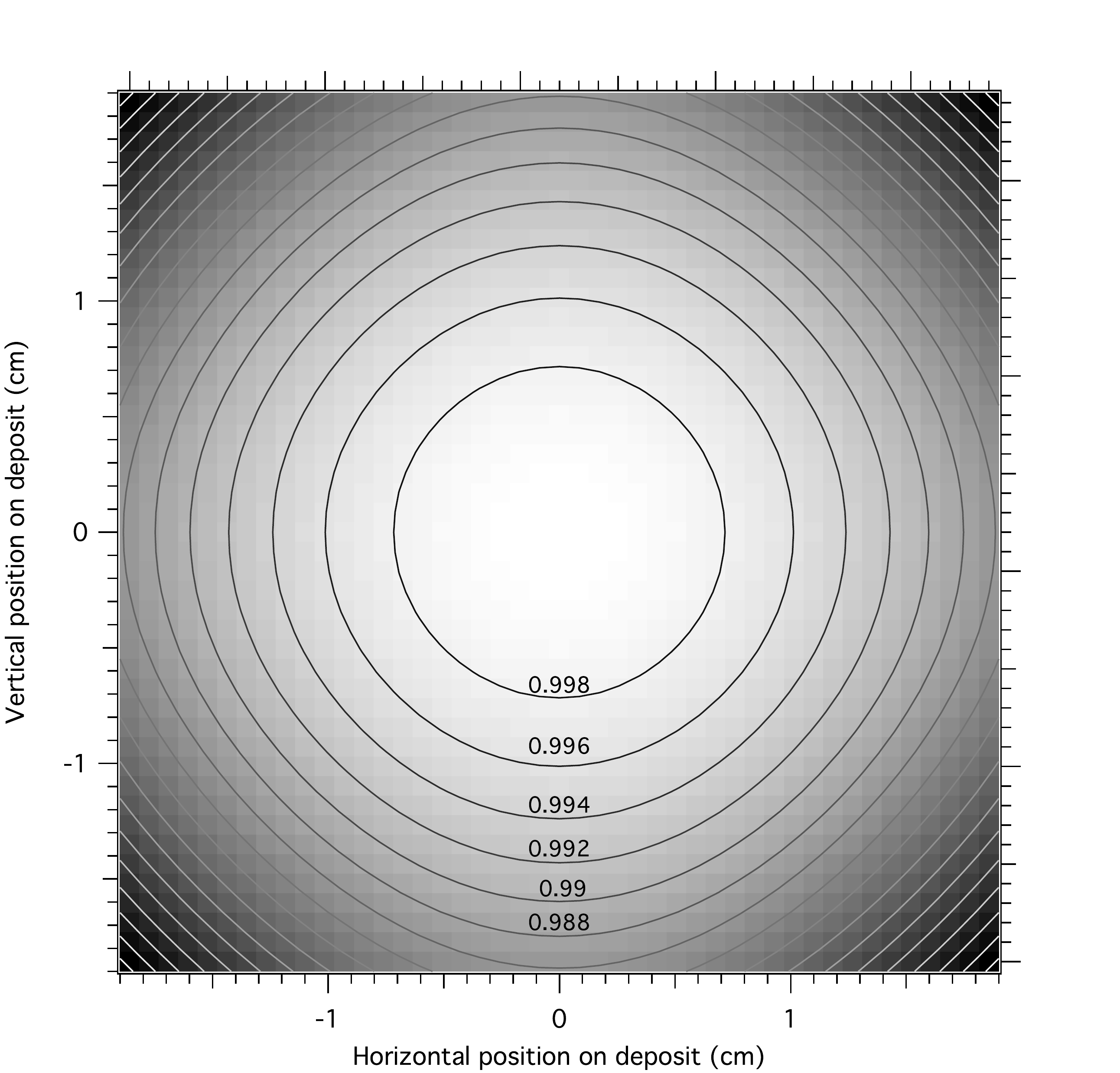}
  \caption{Solid angle as a function of position on the flux monitor deposit.}
  \label{fig:FMOmegaMap}
\end{figure}

\begin{figure}
	\centering
\includegraphics[width=\textwidth]{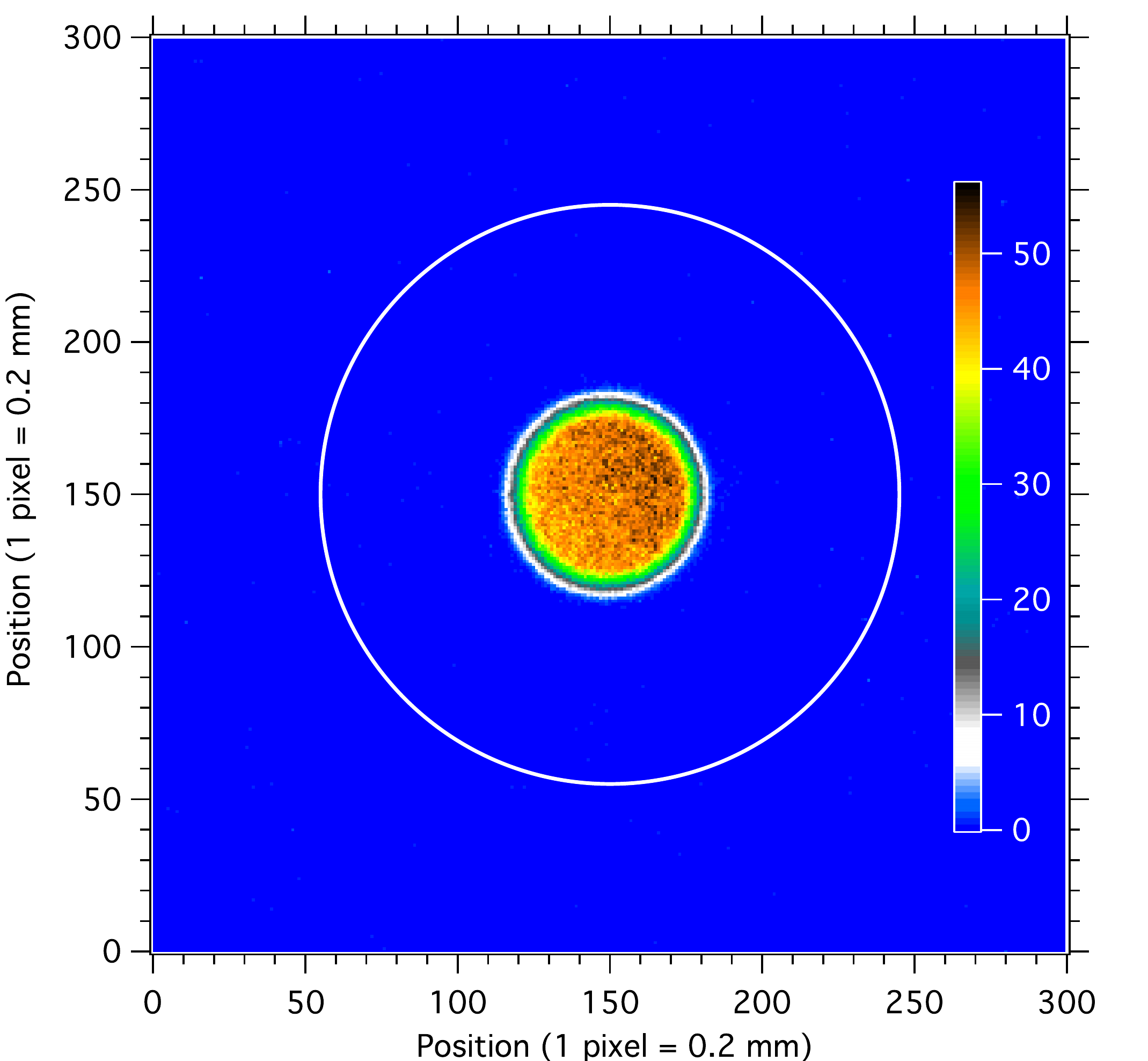}
	\caption{\label{fig:FMBeamImgs}Beam images were acquired at the flux monitor deposit location with C1 = 15\,mm and C2 = 7.2, 8.38, and 10.5\,mm.  The image for the largest collimation of C2 = 10.5\,mm is shown here. The white circle corresponds to the edge of the active area of the ${}^{6}$Li deposit.}
\end{figure}

The neutron flux monitor is subject to corrections for beam solid angle, scattering effects, and the effect of the deposit thickness on neutron absorption.  The response of the monitor to the extended beam used for the calibration must be corrected to the response of a pointlike beam.  This is accomplished with the Dy deposit irradiation method used to determine $\bar{\Omega}_{\rm{AG}}$.  The design of the detector is such that the solid angle to points on the deposit falls off very slowly about the center (figure \ref{fig:FMOmegaMap}).  A beam image of the largest of three collimations is shown in figure \ref{fig:FMBeamImgs}, and the average solid angle ($\bar{\Omega}_{\rm{FM}}$) is found using equation~\ref{eqn:OmegaCalc}.  The average solid angles and the corrections to the flux monitor efficiency are found in table~\ref{tab:FMOmegas}.

The beam extent also leads to a sampling of the deposit density $\bar{\rho} = \int\int \rho(x,y)dxdy$, which must be corrected by a factor of $\rho(0,0)\over{\bar{\rho}}$ in order to establish the idealized efficiency $\epsilon_{0}(0,0)$.  Again using the data provided by the beam images, corrections are established for each beam collimation and given in the last column of table~\ref{tab:FMOmegas}.

\begin{table}
\caption{\label{tab:FMOmegas}Corrections for the flux monitor. The first column gives the three calibration configurations; the second column is the solid angle; the third column gives the correction due to solid angle; and the last column gives the correction due to the beam sampling of the deposit density. The uncertainty arises from the variation seen across multiple Dy images.}
\centering
\begin{tabular}{cccc}
\hline\hline\noalign{\smallskip}
Collimation & $\bar{\Omega}$ (units of $4\pi$) & $\bar{\Omega}_{\rm{FM}}/\Omega_{\rm{FM}}(0,0)$	&	$\rho(0,0)/{\bar{\rho}}$ \\
\noalign{\smallskip}\hline\noalign{\smallskip}
15/7		& 0.00420477	& 1.00043 $\pm$ 0.00016		& 1.00011 $\pm$ 0.00001	\\
15/8		& 0.00420394	& 1.00062 $\pm$ 0.00010		& 1.00015 $\pm$ 0.00001	\\
15/10	& 0.00420336	& 1.00076 $\pm$ 0.00020		& 1.00022 $\pm$ 0.00001	\\
\noalign{\smallskip}\hline\hline
\end{tabular}
\end{table}

Neutrons interact with the silicon substrate of the ${}^{6}$Li deposit as well.  The downstream neutron rate is reduced due to neutron capture and scatter on silicon.  Some neutrons backscatter in the silicon and pass through the deposit again.  The path lengths of these neutrons back through the deposit will be, at a minimum, the same as neutrons initially passing through the deposit.  Thus, backscattered neutrons have an enhanced probability of capture within the deposit, adding a false signal that is highly dependent on the scattering geometry.  In the early 1990s, measurements of this effect were performed on a thermal beam by stacking blank Si wafers behind the deposit and the observed rate is measured as a function of stacked wafers~\cite{Scott95,Nico05}.  It was found that the scattering was larger than one would anticipate if the only scattering was from Bragg scattering and incoherent scattering.  A plausible explanation is that the scattering is occurring at the surfaces of the wafer, where neutrons may scatter from defects or damage.  Only the evaporation surface of the wafer is mirror-polished, making it the most likely scattering candidate.

\begin{figure}[htbp]
  \centering
  \includegraphics[width=\textwidth]{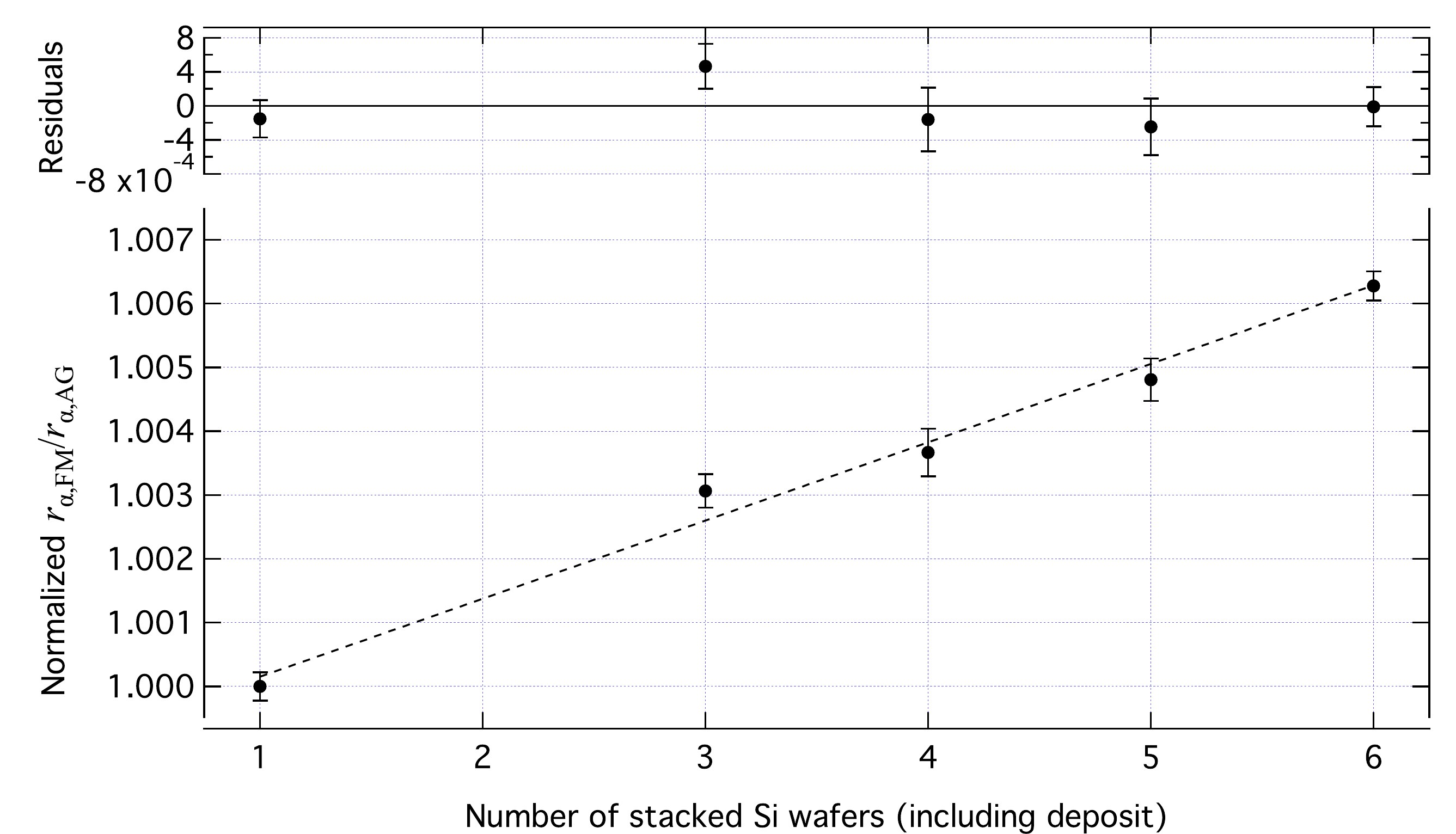}
  \caption{Measured ratio $\frac{r_{\alpha,\rm{FM}}}{r_{\alpha,\rm{AG}}}$ as a function of the number of Si wafers stacked behind the monitor deposit. The dashed line is a linear fit, and the residuals are shown in the upper plot. The error bars represent statistical uncertainty only.}
  \label{fig:AGFMEnhancement}
\end{figure}

The ${}^{6}$Li deposit in the neutron monitor is oriented such that the neutron beam strikes the ${}^{6}$Li layer then the Si backing.  The true rate of neutrons incident on the monitor $R_{n}$ differs from the observed rate of neutrons measured by the Alpha-Gamma device due to neutron absorption and scattering in the Si backing and neutron absorption in the ${}^{6}$Li layer.  The true rate of alphas and tritons in the monitor $r_{\alpha,t}$ is perturbed by backscattered neutrons increasing the signal rate.  The neutron loss due to interaction in a silicon wafer $\eta_{\rm{Si}}$ is given by

\begin{equation}
\eta_{Si} = \epsilon_{\rm{abs}} + \epsilon_{\rm{scatter}},
\end{equation}

\noindent where $\epsilon_{\rm{scatter}}$ is the probability of scatter in the silicon and $\epsilon_{\rm{abs}} = 0.0009$ is the silicon absorption probability~\cite{Sears1992,NCNR2016}.  With the assumption of isotropic scattering, neutrons that backscatter in the silicon ($\frac{\epsilon_{\rm{scatter}}}{2}$) have an average path length $f$ back through the ${}^{6}$Li.  In an ancillary experiment, the approximately 100\,$\mu$g/cm${}^{2}$ B targets were placed in the flux monitor and the Alpha-Gamma device.  The rate of alpha particles measured in the flux monitor $r_{\alpha,\rm{FM}}$ and the rate of alpha particles measured in the Alpha-Gamma $r_{\alpha,\rm{AG}}$ were recorded as a function of the number of wafers $i$ (including the target wafer) that were stacked behind the flux monitor target.  The ratio of the two observed rates as a function of number of stacked wafers is given by

\begin{equation}
\frac{r_{\alpha,\rm{FM}}}{r_{\alpha,\rm{AG}}}\left(i\right) = a\frac{\left(1+\frac{f(i)}{2}i\epsilon_{\rm{scatter}}\right)}{1-i\left(\epsilon_{\rm{scatter}} + \epsilon_{\rm{abs}}\right)},
\end{equation}

\noindent where $a$ is an arbitrary scale parameter. Using the known geometry of the silicon deposits, we calculate $f(i)$.  The measured $\frac{r_{\alpha,\rm{FM}}}{r_{\alpha,\rm{AG}}}\left(i\right)$ is shown in figure \ref{fig:AGFMEnhancement}.  A fit to the data yields $\epsilon_{scatter} = (9.2 \pm 1.8)\times{10}^{-5}$.  This corresponds to a $(2.2 \pm 0.4)\times{10}^{-4}$ enhancement in the flux monitor rate (correction of $0.99978 \pm 0.00004$ to $\epsilon$) and a total neutron attenuation of $\epsilon_{scatter} + \epsilon_{abs} = (1.00 \pm 0.02)\times{10}^{-3}$ (correction of $0.99900 \pm 0.00002$ to $\epsilon$).

Approximately 1\,\% of the beam is absorbed in the ${}^{6}$Li target.  To first order, the absorbed fraction is calculated from the measured ${}^{6}$Li target density, the ENDF/B-VII thermal neutron cross section, and the measured wavelength of the beam

\begin{equation}
\phi_{\rm{abs}} = 1-e^{-\left({{N_{A}}\over{A}}\rho(0,0)\sigma_{0}\frac{\lambda_{\rm{mono}}}{\lambda_{0}}\right)} = 0.01016 \pm 0.00003,
\end{equation}

\noindent and the correction to $\epsilon$ is given by $1-\phi_{\rm{abs}} = 0.98984 \pm 0.00003$.  The absorbed fraction of neutrons in an infinitely thin ${}^{6}$Li target is given by

\begin{equation}
\phi_{\rm{ideal}} = {{N_{A}}\over{A}}\rho(0,0)\sigma_{0}\frac{\lambda_{\rm{mono}}}{\lambda_{0}},
\end{equation}

\noindent and therefore the measured efficiency must be corrected by $\frac{\phi_{\rm{ideal}}}{\phi_{\rm{abs}}} = 1.00512 \pm 0.00001$ in order to convert to $\epsilon_{0}(0,0)$.  From these first order corrections, we follow the procedure outlined in the next section to determine $\epsilon_{0}(0,0)$.  From equation \ref{eqn:e0calc}, we have

\begin{equation}
\rho(0,0)\sigma_{0} = \frac{\epsilon_{0}(0,0)}{\Omega_{\rm{FM}}(0,0)}\frac{A}{N_{A}}.
\end{equation}

\noindent Thus, the measured $\epsilon_{0}(0,0)$ can be used to determine $\rho(0,0)\sigma_{0}$.  This process can be repeated recursively, quickly converging on the true value of $\rho(0,0)\sigma_{0}$.  The final corrections used are $\phi_{\rm{abs}} = 0.989846 \pm 0.000012$ and $\frac{\phi_{\rm{ideal}}}{\phi_{\rm{\rm{abs}}}} = 1.005111 \pm 0.000006$.

Imperfect alignment of the target deposit with respect to the beam direction will lead to an increased neutron path length through the deposit and thus, an enhancement to the observed count rate.  The tilt of the monitor was checked by measuring the center of an alignment target at the deposit position and the downstream flange.  The calculated positions are used to find the tilt angle $\theta = (0.38 \pm 0.02)^{\circ}$.  The path length enhancement is given by $\cos\theta = 0.999978 \pm 0.000002$.

\section{Results}
\label{sec:results}

\begin{figure}[htbp]
  \centering
  \includegraphics[width=\textwidth]{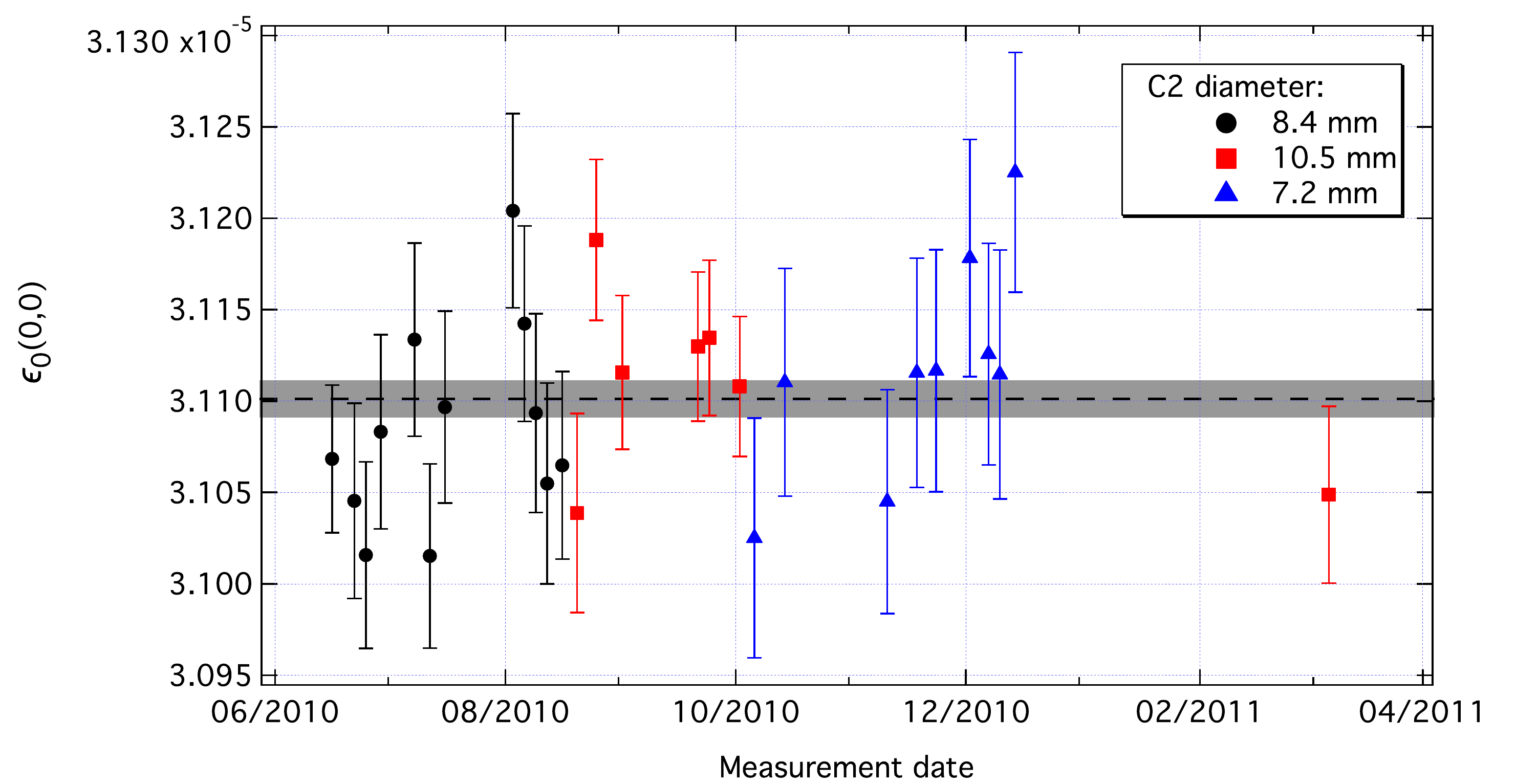}
  \caption{$\epsilon_{0}(0,0)$ measurements for each of the three collimations. The dashed line is the weighted average of the values, and the gray band represents the $\pm 1$-$\sigma$ region. The error bars represent statistical uncertainty only.}
  \label{fig:e0plot}
\end{figure}

Each measurement of $\epsilon_{0}(0,0)$ is found by taking each measurement of $\frac{r_{\alpha,t}}{r_{\gamma,\rm{thick}}}\frac{r_{\gamma,\rm{thin}}}{r_{\alpha,\rm{thin}}}$ and applying the appropriate corrections based on the collimation used to accumulate the data

\begin{equation}
\left.\epsilon_{0}(0,0)\right|_{\rm{C2 = x}} = \left.\left(\frac{r_{\alpha,t}}{r_{\gamma,\rm{thick}}}\frac{r_{\gamma,\rm{thin}}}{r_{\alpha,\rm{thin}}}\right)\right|_{\rm{C2 = x}}\Omega_{\rm{AG}}(0,0)\frac{\lambda_{0}}{\lambda_{\rm{mono}}}\prod_{j=1}^{11}\sqrt{c_{j}^{\rm{T,x}}\times{c_{j}^{\rm{B,x}}}},
\end{equation}

\noindent where $c_{j}^{\rm{T,x}}$ and $c_{j}^{\rm{B,x}}$ are the corrections assigned to the top and bottom detector for downstream collimation x. The 27 measurements of $\epsilon_{0}(0,0)$ and their statistical uncertainty are plotted in figure \ref{fig:e0plot}.

A weighted fit to the data finds $\epsilon_{0}(0,0) = (3.1095 \pm 0.0010) \times{10}^{-5}$, with a $\chi^{2}$/d.o.f. of 1.009.  Each uncertainty was generated by allowing the corresponding correction to vary by its uncertainty with a Gaussian-weighted randomizer.  Each throw of the uncertainty creates a new set of $\epsilon_{0}(0,0)$, which are then fit to a constant with weighting provided by the statistical uncertainty on each point.  By repeating the process many times, the uncertainty can be extracted from the standard deviation in the weighted fit results.  The full uncertainty budget for the experiment is shown in table \ref{tab:THETable}.  The total uncertainty of the neutron flux monitor efficiency is $5.8\times{10}^{-4}$, yielding a final value of $\epsilon_{0}(0,0) = (3.1095 \pm 0.0018)\times{10}^{-5}$.

\begin{table}
\caption{\label{tab:THETable}Systematic effects in the measurement of the flux monitor efficiency and the relative uncertainties. While these numbers were determined specifically for  the nominal 40\,$\mu$g/cm${}^{2}$ deposit, the values are very representative of the lighter deposits. The uncertainties are listed in order of the largest to smallest contribution.}
\centering
\begin{tabular}{lrc}
\hline\hline\noalign{\smallskip}
Systematic effect 										& Relative uncertainty	& 	Section			\\
\noalign{\smallskip}\hline\noalign{\smallskip}
$\alpha$-source solid angle in Alpha-Gamma device			 	& $2.8\times{10}^{-4}$	&	\ref{sec:alphasys}	\\
$\gamma$-ray attenuation by thick target 						& $2.4\times{10}^{-4}$	&	\ref{sec:thicksys}	\\
Neutron beam wavelength 								& $2.3\times{10}^{-4}$	&	\ref{sec:lambdasys}	\\
$\gamma$-ray attenuation by the thin target 					& $1.2\times{10}^{-4}$	&	\ref{sec:thinsyst}	\\
$\lambda/2$ component in beam							& $9.8\times{10}^{-5}$	&	\ref{sec:lambdasys}	\\
$\gamma$-ray signal from absorption in thin target substrate 		& $9.1\times{10}^{-5}$	&	\ref{sec:thinsyst}	\\
Dead time (all sources)									& $8.6\times{10}^{-5}$	&	\ref{sec:thicksys}	\\
Flux monitor enhancement from neutron backscatter 			& $4.0\times{10}^{-5}$	&	\ref{sec:FMcorr}	\\
Alpha-Gamma beam spot solid angle to alpha detector 			& $2.8\times{10}^{-5}$	&	\ref{sec:thinsyst}	\\
Neutron loss in FM deposit substrate 						& $1.8\times{10}^{-5}$	&	\ref{sec:FMcorr}	\\
Neutron absorption by ${}^{6}$Li 							& $1.2\times{10}^{-5}$	&	\ref{sec:FMcorr}	\\

Self-shielding of ${}^{6}$Li deposit 							& $<1\times{10}^{-5}$	&	\ref{sec:FMcorr}	\\
Flux monitor beam spot solid angle 							& $<1\times{10}^{-5}$	&	\ref{sec:FMcorr}	\\
Thick target loss from $^{10}$B(n,$\gamma$) reaction			& $<1\times{10}^{-5}$	&	\ref{sec:thicksys}	\\
Flux monitor misalignment 								& $<1\times{10}^{-5}$	&	\ref{sec:FMcorr}	\\
Surface scatter from B${}_{4}$C 							& $<1\times{10}^{-5}$	&	\ref{sec:thicksys}	\\
\noalign{\smallskip}\hline\noalign{\smallskip}
Neutron counting statistics 								& $3.2\times{10}^{-4}$	&	\ref{sec:results}		\\
\noalign{\smallskip}\hline\noalign{\smallskip}
Total relative uncertainty									& $5.8\times{10}^{-4}$	&					\\
\noalign{\smallskip}\hline\hline
\end{tabular}
\end{table}

\subsection{Prospects for improvement}

The precision achieved in this measurement is sufficient to perform improved measurements of several other quantities that use absolute neutron counting, such as the neutron lifetime and some standard neutron cross sections. There exist, however, avenues for further reduction of some of the larger sources of uncertainty that are listed in Table~\ref{tab:THETable}. The largest source of uncertainty arises from the solid angle calibration using the alpha source, and improvement in precision could come from a redesign of the alpha particle detection system in the Alpha-Gamma apparatus. Instead of a single silicon detector, four detectors could be arranged in the same manner as the flux monitor. This would give two beneficial effects. The viewing angle of the deposit to the silicon could be adjusted to better suppress systematics associated with the spot size. Also, the apertures that define the solid angle from the deposit would be rigidly mounted to the deposit. That would eliminate sensitivity to any slight movements in relative positions of the deposit, apertures, and detectors. Additionally, one could perform contact metrology on the rigid structure to precisely determine dimensions, which was impossible with the existing vacuum-supported kinematic setup. It should be possible to implement these changes without losing the ability to quickly change target states.

The next largest systematic effect is the attenuation of gamma rays in the thick target. This could be nearly eliminated by arranging four gamma detectors in front of the deposit position, where the escaping gamma rays would no longer have to traverse the longer distance through the thick target and the substrate of the thin target. Another benefit is that one can suppress substrate and target scattering effects. Because orientation of the HPGe crystal is only significant for path-length-dependent uncertainties, suppressing any scattering effects also suppresses the dependence of the experiment on the orientation of the detectors. Using four detectors instead of two has the obvious benefit of increasing the data acquisition rate of the thin target measurements.

The wavelength of the neutron beam also has large uncertainty associated with it, but this is largely due to the inflation of the error bar due to the non-statistical spread in the wavelength measurements. This spread was due to one group of measurements. Those measurements were not investigated at the time because of their comparatively small impact on the overall uncertainty. Although we don't know the reason for the excursion for those data points, we are fairly confident that it could be understood and eliminated with a focused effort, thus reducing the error for the wavelength measurement.

By eliminating those three largest systematic effects and reducing statistical uncertainty through both increased run time and increased solid angles, one could reasonably anticipate an improvement in the overall uncertainty of about a factor of 4.

\section{Summary and Applications}
\label{sec:summary}

A device for absolute neutron flux measurements was constructed based on a novel cross calibration procedure that ultimately couples the measurement to simple geometry and chronometry. It uses measured counting rates and metrology and does not rely upon cross sections or branching ratios as inputs. The relative uncertainties achieved with this technique have demonstrated sub-\(0.1\%\) precision. Using this device, we produced an absolute efficiency for a flux monitor that is in agreement with a previous determination. That efficiency has been used to improve the precision of the neutron lifetime value from an experiment that used the identical flux monitor in its measurement~\cite{Yue2013}. With this method established, one can use the technique to perform other measurements that require knowledge of the absolute neutron flux; we briefly mention two applications.

\subsection{Re-calibration of NBS-1}
\label{sec:recal}

The Alpha-Gamma device can be used as part of a recalibration of NBS-1, a Ra-Be photo-neutron source that is the cornerstone of neutron dosimetry in the US.  NBS-1 is used as a comparison standard for calibrating neutron sources and neutron dosimeters and also in establishing standard fast and thermal neutron fields. As such, the uncertainty in the calibration of NBS-1 sets a lower limit for the uncertainty in any service derived from its neutron emission rate~\cite{Adams04}. In its first calibration, the thermal neutron density was  determined from the activity induced in thin indium and magnesium foils as a function of distance from the source when both the source and foils were under water. The second calibration involved the capture of neutrons by a surrounding manganese sulfate (MnSO$_4$) bath followed by a measurement of the activity of the $^{56}$Mn. A third method involved a relative comparison to an antimony-beryllium source that had been calibrated absolutely in a heavy-water manganese sulfate bath. With these measurements, the overall uncertainty was determined to be $\pm 1$\,\%. Subsequent to those measurements, the neutron emission rate was checked against the number of $^{252}$Cf neutrons that are emitted per fission, known as $\bar{\nu}$~\cite{McGarry1988}, and the uncertainty was reduced to its current value of $\pm 0.85$\,\%.

This approach to the absolute calibration of NBS-1 is sufficiently difficult that it has not been done for over forty years, but the Alpha-Gamma technique offers a novel method to check the absolute neutron emission rate and also improve the precision. Neutron source emission rates are measured at NIST using a MnSO$_4$ bath technique wherein the neutron-induced activity of the Mn in a large bath of MnSO$_4$ dissolved in water is compared between an unknown source and NBS-1. The known emission rate of NBS-1 allows the determination of the emission rate of the unknown source from the ratio of induced activities.  The recalibration of NBS-1 requires a small transfer neutron source that is similar to NBS-1, a small, portable manganese bath, and the flux monitor after calibration by the Alpha-Gamma device.  The portable bath would be calibrated by installing it downstream of the flux monitor on a monochromatic neutron beam and simultaneously measuring the flux monitor response and the induced Mn activity.  Subsequently, the transfer source emission rate would be measured in the calibrated portable bath, and finally, the induced Mn activity is compared between NBS-1 and the transfer source in the large bath, thus establishing a new emission rate for NBS-1. We estimate that the overall uncertainty on the neutron emission rate of NBS-1 could be reduced by as much as a factor of 3.
\subsection{Standard neutron cross sections}
\label{sec:xsections}

A significant advantage of using the Alpha-Gamma method to establish the counting efficiency of a device such as the flux monitor is that the determination is independent of parameters of the monitor, such as its solid angle, the mass of the target deposit, and neutron cross section of the deposit material. From equation~\ref{eqn:e0calc}, one can see that if in addition to measuring the efficiency, one also knew the solid angle and the deposit mass, one could extract the neutron cross section of deposit material. In this work, \Ls\ was used as the neutron absorbing material of the deposit. The ${}^{6}$Li(n,t)${}^{4}$He cross section is important in nuclear physics, astrophysics, applications of nuclear technology, and also as a standard in determining other cross sections. As we know both the mass of the \Ls\ and the solid angle of the flux monitor with good precision, it is possible to determine the neutron cross section in a novel way. For such a measurement, the uncertainty in the cross section would be dominated by the determination of the areal density of the ${}^{6}$Li target deposit, but it is not a fundamentally limiting systematic effect.

This method is not limited to \Ls\ but can be extended to other isotopes that can be fabricated into well-characterized deposits and whose reaction products lends themselves to efficient detection in the flux monitor. It offers a path for significant improvement to several low-energy standard neutron cross sections. Two of the more important reaction cross sections that could be measured are ${}^{10}$B(n,$\alpha$)${}^{7}$Li and ${}^{235}$U(n,f). Thin samples of both of these isotopes can be produced and characterized for their mass and deposition profile. They would be placed in the monochromatic neutron beam and the absolute neutron flux measured in largely the same manner as with the \Ls\ deposit. Minor modifications in the technique may be required depending upon the reaction products.

\ack

We thank J. R. Stoup of NIST for performing the dimensional metrology on the neutron flux monitor rig. We also thank A. D. Carlson and H. P. Mumm of NIST  and F. E. Wietfeldt of Tulane University for their long interest regarding this work. The authors acknowledge the support of the National Institute of Standards and Technology, U.S. Department of Commerce, in providing the neutron research and chemistry facilities used in this work, and they gratefully acknowledge the support of NIST (U.S. Department of Commerce), the U.S. Department of Energy Office of Nuclear Physics (Grants No. DE-SC0005925 and No. DE-FG02- 03ER41258). E. Anderson and W. M. Snow acknowledge support from the U.S. National Science Foundation grant PHY-1306942,  a NIST Precision Measurement Grant 70NANB14H299, and the Indiana University Center for Spacetime Symmetries.

\section*{References}
\providecommand{\newblock}{}


\begin{thebibliography}{10}
\expandafter\ifx\csname url\endcsname\relax
  \def\url#1{{\tt #1}}\fi
\expandafter\ifx\csname urlprefix\endcsname\relax\def\urlprefix{URL }\fi
\providecommand{\eprint}[2][]{\url{#2}}

\bibitem{Shultis2007}
Shultis J~K and Faw R~E 2007 {\em {Fundamentals of Nuclear Science and
  Engineering}\/} 2nd ed (CRC Press)

\bibitem{Harling1995}
Harling O~K, Moulin D, Chabeuf J~M and Solares G~R 1995 {\em Nucl. Instrum.
  Meth. B\/} {\bf 101} 464

\bibitem{Raaijmakers1996}
Raaijmakers C~P~J, Nottelman E~L, Konijnenberg M~W and Mijnheer B~J 1996 {\em
  Phys. Med. Biol.\/} {\bf 41} 2789

\bibitem{Tanner1999}
Tanner V, Auterinen I, Helin J, Kosunen A and Savolainen S 1999 {\em Nucl.
  Instrum. Meth. A\/} {\bf 422} 101 -- 105

\bibitem{Beckurts1964}
Beckurts K~H and Wirtz K 1964 {\em {Neutron Physics}\/} 2nd ed
  (Springer-Verlag)

\bibitem{Knoll2000}
Knoll G~F 1989 {\em {Radiation Detection and Measurement}\/} 2nd ed (Wiley, New
  York)

\bibitem{Williams2011}
Williams J~G and Gilliam D~M 2011 {\em Metrologia\/} {\bf 48} S254

\bibitem{Greenberg2011}
Greenberg R~R, Bode P and Fernandes E~A~D~N 2011 {\em Spectrochimica Acta Part
  B: Atomic Spectroscopy\/} {\bf 66} 193 -- 241

\bibitem{deSoete1972}
Soete D~D, Gijbels R and Hoste J 1972 {\em Neutron Activation Analysis\/} (John
  Wiley and Sons: New York, NY.)

\bibitem{Fraser1995}
FRASER G {1995} {\em {Neutrons and their Applications, International
  Conference}\/} ({\em {Proceedings of the Society of Photo-Optical
  Instrumentation Engineers (SPIE)}\/} vol {2339}) ed {Vourvopoulos, G and
  Paradellis, T} pp {287--301}

\bibitem{Brenizer2013}
Brenizer J~S 2013 {\em Physics Procedia\/} {\bf 43} 10

\bibitem{Nolte2015}
Nolte R, B\"{o}ttger R, Chen J, Harano H and Thomas D~J 2015 {\em Metrologia\/}
  {\bf 52} 06011

\bibitem{Gressier2014}
Gressier V, Bonaldi A~C, Dewey M~S, Gilliam D~M, Harano H, Masuda A, Matsumoto
  T, Moiseev N, Nico J~S, Nolte R, Oberstedt S, Roberts N~J, R\"{o}ttger S and
  Thomas D~J 2014 {\em Metrologia\/} {\bf 51} 06009

\bibitem{Thomas2011}
Thomas D~J, Nolte R and Gressier V 2011 {\em Metrologia\/} {\bf 48}

\bibitem{Wietfeldt2011}
Wietfeldt F~E and Greene G~L 2011 {\em Rev. Mod. Phys.\/} {\bf 83} 1173--1192

\bibitem{Byrne96}
Byrne J, Dawber P~G, Habeck C~G, Smidt S~J, Spain J~A and Williams A~P 1996
  {\em Europhys. Lett.\/} {\bf 33} 187--192

\bibitem{Dewey2003}
Dewey M~S, Gilliam D~M, Nico J~S, Wietfeldt F~E, Fei X, Snow W~M, Greene G~L,
  Pauwels J, Eykens R, Lamberty A and VanGestel J 2003 {\em Phys. Rev. Lett.\/}
  {\bf 91} 152302

\bibitem{Nico05}
Nico J~S, Dewey M~S, Gilliam D~M, Wietfeldt F~E, Fei X, Snow W~M, Greene G~L,
  Pauwels J, Eykens R, Lamberty A, VanGestel J and Scott R~D 2005 {\em Phys.
  Rev. C\/} {\bf 71} 055502

\bibitem{Dewey08}
Dewey M~S, Gilliam D~M, Nico J~S, Greene G~L, Laptev A~B and Yue A~T 2008 {\em
  Proceedings, 13th ASTM-EWGRD Reactor Dosimetry Symposium\/}

\bibitem{Adams04}
Adams J~M 2004 {\em Nucl. Instrum. Meth. B\/} {\bf 213} 218--222

\bibitem{Gilliam08}
Gilliam D~M, Yue A~T and Dewey M~S 2008 {\em Nucl. Instrum. Meth. A\/} {\bf
  590} 181--184

\bibitem{Seltzer2011}
Seltzer S~M, Bartlett D~T, Burns D~T, Dietze G, Menzel H~G, Paretzke H~G and
  Wambersie A 2011 {\em Journal of the ICRU\/} {\bf 11} 1

\bibitem{Lamaze88}
Lamaze G~P, Gilliam D~M and Williams A~P 1988 {\em Journal of Radioanalytical
  and Nuclear Chemistry, Articles\/} {\bf 123} 551--559

\bibitem{Robertson86}
Robertson R~G~H and Koehler P~E 1986 {\em Nucl. Instrum. Meth. A\/} {\bf 251}
  307

\bibitem{Wietfeldt09}
Wietfeldt F~E 2009 {Absolute Thermal Neutron Fluence Measurement Using
  ${}^{3}$He Gas Scintillation} unpublished

\bibitem{Nagakura2017}
Nagakura N {\em et~al.\/} 2017 {\em PoS\/} {\bf INPC2016} 191

\bibitem{Richardson93}
Richardson J~M 1993 {\em Accurate Determination of Thermal Neutron Flux: A
  Critical Step in the Measurement of the Neutron Lifetime\/} Ph.D. thesis
  Harvard University

\bibitem{Chowdhuri00}
Chowdhuri Z 2000 {\em Absolute Neutron Measurements in Neutron Decay\/} Ph.D.
  thesis Indiana University

\bibitem{Chowdhuri03}
Chowdhuri Z, Hansen G~L, Jane V, Keith C~D, Lozowski W~M, Snow W~M, Dewey M~S,
  Gilliam D~M, Greene G~L, Nico J~S and Thompson A~K 2003 {\em Rev. Sci.
  Instrum.\/} {\bf 74} 4280--4293

\bibitem{Hansen04}
Hansen G~L 2004 {\em A Radiometric Measurement of Neutron Flux in a Liquid
  ${}^{3}$He Target\/} Ph.D. thesis Indiana University

\bibitem{Gilliam89}
Gilliam D~M, Greene G~L and Lamaze G~P 1989 {\em Nucl. Instrum. Meth. A\/} {\bf
  284} 220--222

\bibitem{Denecke99}
Denecke B, Eykens R, Pauwels J, Robouch P, Gilliam D~M, Hodge P, Hutchinson
  J~M~R and Nico J~S 1999 {\em Nuclear Instruments \& Methods in Physics
  Research, Section A\/} {\bf 438} 124--130

\bibitem{Deruytter67}
Deruytter A~J and Pelfer P 1967 {\em Journal of Nuclear Energy\/} {\bf 21}
  833--845

\bibitem{Stelts79}
Stelts M~L, Chrien R~E, Goldhaber M and Kenny M~J 1979 {\em Physical Review
  C\/} {\bf 19} 1159--1167

\bibitem{Williams89}
Williams A~P 1990 {\em The Determination of the Neutron Lifetime by Trapping
  Decay Protons\/} Ph.D. thesis University of Sussex

\bibitem{Yue2011}
Yue A~T 2011 {\em Progress Towards a Redetermination of the Neutron Lifetime
  Through the Absolute Determination of Neutron Flux\/} Ph.D. thesis University
  of Tennessee

\bibitem{Pauwels95}
Pauwels J, Scott R~D, Eykens R, Robouch P, Gestel J~V, Verdonck J, Gilliam D~M
  and Greene G~L 1995 {\em Nuclear Instruments \& Methods in Physics Research,
  Section A\/} {\bf 362} 104--111

\bibitem{Scott95}
Scott R~D, D'hondt P, Eykens R, Robouch P, Reher D~F~G, Sibbens G, Pauwels J
  and Gilliam D~M 1995 {\em Nuclear Instruments \& Methods in Physics Research,
  Section A\/} {\bf 362} 151--159

\bibitem{ENDF7}
Carlson A~D {\em et~al.\/} 2009 {\em Nucl. Data Sheets\/} {\bf 110} 3215--3324

\bibitem{Rush2011}
Rush J~J and Cappelletti R~L 2011 {The NIST Center for Neutron Research} {NIST
  SP1120}

\bibitem{NicoJRES05}
Nico J~S, Arif M, Dewey M~S, Gentile T~R, Gilliam D~M, Huffman P~R, Jacobson
  D~L and Thompson A~K 2005 {\em J. Res. Natl. Inst. Stand. Technol.\/} {\bf
  110} 137--144

\bibitem{Mohr06}
Mohr P~J, Taylor B~N and Newell D~B 2006 {\em Rev. Mod. Phys.\/} {\bf 80}
  633--730

\bibitem{McStas2014}
Willendrup P, Farhi E, Knudsen E, Filges U and Lefmann K 2014 {\em J. of
  Neutron Research\/} {\bf 17} 35

\bibitem{Coakley2003}
Coakley K~J, Chowdhuri Z, Snow W~M, Richardson J~M and Dewey M~S 2003 {\em
  Meas. Sci. Technol.\/} {\bf 14} 131--139

\bibitem{Moon1937}
Moon P~B 1937 {\em J. Sci. Instrum.\/} {\bf 14} 189

\bibitem{Janssen1994}
Janssen H and Shoenfeld E 1994 {\em Nucl. Instrum. Meth. A\/} {\bf 339} 137

\bibitem{XCOM}
Berger M~J, Hubbell J~H, Seltzer S~M, Chang J, Coursey J~S, Sukumar R, Zucker
  D~S and Olsen K 2010 Xcom: Photon cross section database (version 1.5)
  available online at http://physics.nist.gov/xcom

\bibitem{Sears1992}
Sears V~F 1992 {\em Neutron News\/} {\bf 3} 26--37

\bibitem{NCNR2016}
 2016 Https://www.ncnr.nist.gov/resources/n-lengths/elements/b.html

\bibitem{Anders1969}
Anders O~U 1969 {\em Nucl. Instrum. Meth.\/} {\bf 68} 205

\bibitem{Bartholomew1957}
Bartholomew G~A and Campion P~J 1957 {\em Canadian Journal of Physics\/} {\bf
  35} 1347--1360 (\textit{Preprint} \eprint{https://doi.org/10.1139/p57-147})
  \urlprefix\url{https://doi.org/10.1139/p57-147}

\bibitem{Kok1986}
Kok P~J~J, de~Haas J~B~M, Abrahams K, Postma H and Huiskamp W~J 1986 {\em Z.
  Phys. A\/} {\bf 324} 271

\bibitem{Firestone2016}
Firestone R~B and Revay Z 2016 {\em Physical Review C\/} {\bf 93} 054306 ISSN
  2469-9985 \urlprefix\url{http://link.aps.org/doi/10.1103/PhysRevC.93.054306}

\bibitem{Yue2013}
Yue A~T, Dewey M~S, Gilliam D~M, Greene G~L, Laptev A~B, Nico J~S, Snow W~M and
  Wietfeldt F~E 2013 {\em Phys. Rev. Lett.\/} {\bf 111}(22) 222501

\bibitem{McGarry1988}
McGarry E~D and Boswell E~W 1988 {Neutron Source Strength Calibrations, NBS
  Special Publication 250-18}

\end{thebibliography}
\end{document}